\documentclass[journal]{IEEEtran}

\usepackage{cite}
\usepackage{amsmath}
\usepackage{amssymb}
\usepackage{amsfonts}
\usepackage{cases}
\usepackage{graphicx}
\usepackage{color}
\usepackage{multirow}
\usepackage{booktabs}
\usepackage{relsize}
\usepackage{stfloats}
\usepackage{xcolor}
\usepackage{enumerate}
\usepackage{float}
\usepackage{setspace}
\usepackage{verbatim}
\usepackage{bm}
\usepackage{xfrac}
\usepackage{algpseudocode}
\usepackage{subcaption}
\usepackage{dsfont}
\usepackage{lipsum}

\usepackage{pgfplots}
\pgfplotsset{compat=1.17} 

\newtheorem{remark}{Remark}

\newtheorem{proposition}{Proposition}

\newcommand{\rom}[1]{\uppercase\expandafter{\romannumeral #1\relax}}

\allowdisplaybreaks

\begin{document}
\title{Cognitive-Radio Functionality: A Novel Configuration for STAR-RIS assisted RSMA Networks}

\author{Saeed Ibrahim, Yue Xiao,~\IEEEmembership{Member,~IEEE,} Dimitrios~Tyrovolas,~\IEEEmembership{Member,~IEEE,}\\ Sotiris A. Tegos,~\IEEEmembership{Senior Member,~IEEE}, Panagiotis D. Diamantoulakis,~\IEEEmembership{Senior Member,~IEEE}, \\Zheng Ma,~\IEEEmembership{Member,~IEEE}, ~George K.~Karagiannidis,~\IEEEmembership{Fellow,~IEEE}, and Pinghzi Fan,~\IEEEmembership{Fellow,~IEEE}

\thanks{
Saeed Ibrahim, Yue Xiao, Zheng Ma and Pingzhi Fan are with the School of Information Science and Technology, Southwest Jiaotong University, 610031, Chengdu, China (e-mail: saeedibrahim@my.swjut.edu.cn, alice\_xiaoyue@hotmail.com, zma@swjtu.cn, pzfan@home.swjtu.edu.cn)

D. Tyrovolas, S. A. Tegos, P. D. Diamantoulakis, and G. K. Karagiannidis are with the Department of Electrical and Computer Engineering, Aristotle University of Thessaloniki, 54124 Thessaloniki, Greece (tyrovolas@auth.gr, tegosoti@auth.gr, padiaman@auth.gr, geokarag@auth.gr).}
}

\maketitle
\begin{abstract}
Cognitive radio rate-splitting multiple access (CR-RSMA) has emerged as a promising multiple access framework that can efficiently manage interference and adapt dynamically to heterogeneous quality-of-service (QoS) requirements. To effectively support such demanding access schemes, programmable wireless environments have attracted considerable attention, especially through simultaneously transmitting and reflecting reconfigurable intelligent surfaces (STAR-RISs), which can enable full-space control of signal propagation in asymmetric user deployments. 
In this paper, we propose the cognitive radio (CR) functionality for STAR-RIS-assisted CR-RSMA systems, leveraging the unique capability of the STAR-RIS to combine element and power splitting for adaptive control of transmission and reflection in CR scenarios. Specifically, the proposed CR functionality partitions the STAR-RIS into two regions independently controlling the transmission and reflection of signals, simultaneously ensuring the required QoS for the primary user and enhancing the performance of the secondary user. To accurately characterize the system performance, we derive analytical expressions for the ergodic rate of the secondary user and the outage rate of the primary user under Nakagami-$m$ fading.
Finally, simulation results show that the proposed approach effectively manages interference, guarantees the QoS of the primary user, and significantly improves the throughput of the secondary user, highlighting STAR-RIS as an efficient solution for CR-RSMA-based services.
\end{abstract}
\section{Introduction}

The evolution of wireless networks into increasingly dense, heterogeneous, and service-oriented environments has intensified the demand for access mechanisms that are both spectrally efficient and dynamically adaptable \cite{DINGSurvey2017}. As a wide range of applications imposes diverse quality-of-service (QoS) requirements under constrained spectral resources, traditional schemes based on orthogonality or fixed allocation often fail to meet simultaneous user demands \cite{Mishra2022RSMA}. This growing limitation has raised interest in frameworks that can manage user heterogeneity, coordinate interference, and exploit spectrum opportunities more flexibly \cite{Mishra2022RSMA,LiU2017NOMA}. Consequently, there is a growing need to unify spectral awareness with advanced resource control, in order to enable access strategies that can dynamically accommodate users with varying priorities while adapting to real-time network conditions.

Building on the need for more flexible and interference-aware access mechanisms, recent efforts have focused on transmission strategies capable of supporting user coexistence under spectrum scarcity while satisfying diverse QoS requirements \cite{proceed}. In this direction, cognitive radio rate-splitting multiple access (CR-RSMA) has emerged as a compelling framework that combines the spectrum awareness of cognitive radio (CR) with the layered decoding structure of RSMA \cite{Tegos2022, LiuCRNOMA2022,YUE2023CR-RSMA}. Specifically, CR-RSMA enables network users to dynamically share spectral resources while adapting transmission strategies in response to channel conditions and user-specific constraints. Furthermore, the ability of CR-RSMA to manage asymmetric interference and prioritize access based on real-time network conditions makes it particularly suitable for environments characterized by heterogeneous service demands and variable spatial configurations \cite{YUE2023CR-RSMA}. However, the practical effectiveness of CR-RSMA is determined not only by transmitter-side adaptation or receiver-side decoding, but also by the physical conditions under which signals interact, interfere, and are separated in the network, as its performance remains closely coupled to relative signal strengths, interference dynamics, and the spatial distribution of users \cite{Munochiveyi2021}. Therefore, maintaining robust and reliable operation under dynamic channel conditions requires a reconfigurable wireless infrastructure capable of adapting the propagation environment to support the layered and interference-aware structure of CR-RSMA.

To meet the demands introduced by adaptive access strategies such as CR-RSMA, the concept of programmable wireless environments (PWEs) has emerged as a fundamental approach to controlling the wireless propagation process \cite{liaskosmagazine}. Rather than treating the environment as a passive medium, PWEs embed reconfigurable components in physical space, enabling dynamic manipulation of electromagnetic wave behavior in response to real-time network conditions and user requirements \cite{direnzo2022, tegos2022distribution}. At the core of this paradigm is the development of reconfigurable technologies that allow the wireless medium to be precisely shaped and adapted. In particular, reconfigurable intelligent surfaces (RISs) have gained significant attention as a practical solution for implementing the PWE vision, using arrays of tunable reflecting elements to manipulate the signals impinging upon them in ways that enhance communication performance. As a result, this has made RISs well suited to extend coverage, mitigate interference, and shape the spatial properties of the channel to support flexible and interference-aware transmissions, especially in scenarios where propagation must align with complex and asymmetric service demands \cite{direnzo2022, tegos2022distribution, yuanwei2021RIS}. Additionally, to accommodate diverse deployment scenarios and the increasing complexity of modern access mechanisms, RIS designs have evolved into a variety of architectures, each offering distinct signal control capabilities and structural features \cite{liu2021star,zeris, leris, activeris}. Consequently, supporting CR-RSMA under dynamic and interference-sensitive conditions requires evaluating which RIS architectures can most effectively meet its reconfigurability and signal control requirements.

\subsection{Literature Review \& Motivation}

The integration of RISs into CR systems has attracted considerable attention due to their ability to reshape the wireless propagation environment and facilitate efficient spectrum sharing \cite{YuanJie2021}. This capability has motivated the use of RISs to support CR-inspired multiple access strategies, where interference management and real-time adaptability are critical. In this context, several studies have investigated RIS-supported CR-NOMA systems, demonstrating that RIS can increase secondary throughput \cite{Vu2022}, support high-capacity multi-user access under performance constraints of the primary user (PU) \cite{Weiheng2022}, and improve reliability in short-packet transmissions for latency-sensitive scenarios \cite{xia2024shortpacket}. Collectively, these contributions highlight the potential of RISs to enable stringent interference-limited access, although they remain rooted in fixed power-domain overlays and static decoding hierarchies that limit adaptability to varying interference levels and heterogeneous user requirements \cite{Lv2025MEC}. However, while \cite{Vu2022,Weiheng2022,xia2024shortpacket} confirm the effectiveness of RIS in managing interference in NOMA-based CR systems, they do not provide transferable insights for CR-RSMA, which differs fundamentally from NOMA in both transmission structure and interference management. To explore this direction, the authors in \cite{RSMATsiftsis2023} investigated a RIS-assisted CR-RSMA uplink system and analyzed how the integration of RISs enables reliability improvements of the secondary user (SU) while maintaining PU protection through joint optimization of power allocation, rate splitting (RS), and decoding strategies. Specifically, their results confirm that RSMA outperforms NOMA in supporting secondary transmissions under interference constraints, thereby strengthening the compatibility between RIS and CR-RSMA. However, the proposed model assumes a specific RIS design and does not examine how various RIS architectures or deployment strategies impact system performance. Consequently, while these studies collectively demonstrate that RISs are capable of supporting CR-RSMA, further investigation is needed to identify which RIS architectures are best suited to fulfill the stringent and adaptive requirements of CR-RSMA-driven services.

Given the demanding requirements of CR-RSMA in terms of spatial adaptability and interference-aware transmission, it becomes increasingly important to examine RIS architectures that offer greater flexibility than conventional passive surfaces, which operate solely through reflection and lack the ability to support dynamic multi-user coordination. This motivates the consideration of alternative designs that enable additional control over wave propagation. In this context, simultaneously transmitting and reflecting reconfigurable intelligent surfaces (STAR-RISs) have emerged as a compelling solution, as they allow electromagnetic waves to be both transmitted and reflected through different sides of the surface, thus enabling full-space signal manipulation. These capabilities make STAR-RIS particularly suitable for environments with asymmetric user distributions and directional service requirements \cite{li2024star}. In this direction, several recent works have applied STAR-RIS to CR-NOMA systems, confirming its ability to improve reliability, secrecy, and throughput in interference-constrained settings. For example, the authors in \cite{Wang2023STAROutage} and \cite{Security2024STAR} showed that STAR-RIS improves PU protection while enhancing SU performance, even under fading and eavesdropping conditions, while \cite{IndustrySTAR2024} demonstrated gains in energy efficiency and ergodic rate for STAR-RIS-assisted IoT networks. However, these studies are still focused on NOMA and do not address the layered transmission and adaptive decoding required by RSMA. Thus, to the best of the authors’ knowledge, no prior work has investigated the use of STAR-RIS in CR-RSMA systems, nor proposed a cognitive configuration policy based on closed-form expressions that can dynamically adapt the STAR-RIS surface in response to network conditions.

\subsection{Contribution}
In this paper, we introduce a unique STAR-RIS-enabled electromagnetic functionality, referred to as \textit{CR functionality}, which is uniquely realized by STAR-RISs to support the requirements of CR-RSMA transmission. Specifically, by embedding cognitive principles into the STAR-RIS configuration, the proposed functionality leverages the simultaneous transmission and reflection capabilities of the STAR-RIS to adaptively reshape the wireless environment to ensure PU QoS and improve SU performance. In particular, this is achieved by jointly applying element splitting and power splitting, where the STAR-RIS is segmented into two independent regions tailored to the different requirements of each user, and the proportion of power reflected or transmitted from each region is controlled to modify the channel gains accordingly. As a result, the CR functionality enables, for the first time, performance analysis of CR-RSMA systems in PWEs empowered by STAR-RIS. More specifically, the main contributions of this work are summarized as follows:
\begin{itemize}
    \item We propose the CR functionality for STAR-RIS-assisted CR-RSMA systems, leveraging the unique capability of STAR-RIS to combine element and power splitting for adaptive control of transmission and reflection in CR scenarios.
    \item We establish a CR-RSMA uplink framework and derive  SINR expressions that reflect user priorities and interference conditions.
    \item We derive analytical expressions for the SU's ergodic rate and the PU's outage probability under Nakagami-$m$ fading, and develop a case-based performance framework based on user channel conditions and RSMA parameters.
    \item We validate the theoretical analysis through simulations and demonstrate the impact of STAR-RIS element allocation and RS power control on CR-RSMA system performance.
\end{itemize}

\subsection{Structure}
The remaining of the paper is organized as follows. Section II presents the system model and describes the STAR-RIS-assisted CR-RSMA framework, while Section III provides the analytical performance analysis, deriving closed-form expressions for the ergodic rate and the outage rate of the SU and the PU respectively, while Section IV presents numerical results that validate the analysis and offer insights into the system behavior. Finally, Section V concludes the paper.

\section{Cognitive-Radio Functionality}

\subsection{System Overview}
We consider an uplink communication scenario in a dense urban environment with harsh propagation conditions, where two single-antenna users within a PWE aim to transmit their signals to a single-antenna base station (BS) with complete channel state information. In order to enable reliable communication by effectively managing interference, the network employs the CR-RSMA paradigm which facilitates simultaneous transmissions while handling the varying interference levels and QoS requirements of heterogeneous users. Specifically, in the CR-RSMA framework, a PU is granted prioritized access to network resources to ensure that its QoS requirements are met, while an SU dynamically adapts its transmission strategy to coexist harmoniously with the PU. In addition, by using RSMA, the SU splits its message $x_s$ into two sub-messages, $x_{s_1}$ and $x_{s_2}$, by allocating its transmit power using an RS factor $\alpha$ with $\alpha \in [0,1]$, allowing simultaneous communication with the PU and facilitating interference management. This interplay between the PU and SU ensures that the network achieves a balanced trade-off between interference control and spectral efficiency, providing the basis for reliable communication in complex propagation environments \cite{YUE2023CR-RSMA,10014691}.
\begin{figure}
\centerline{\includegraphics[width=\columnwidth, height=5.5cm]{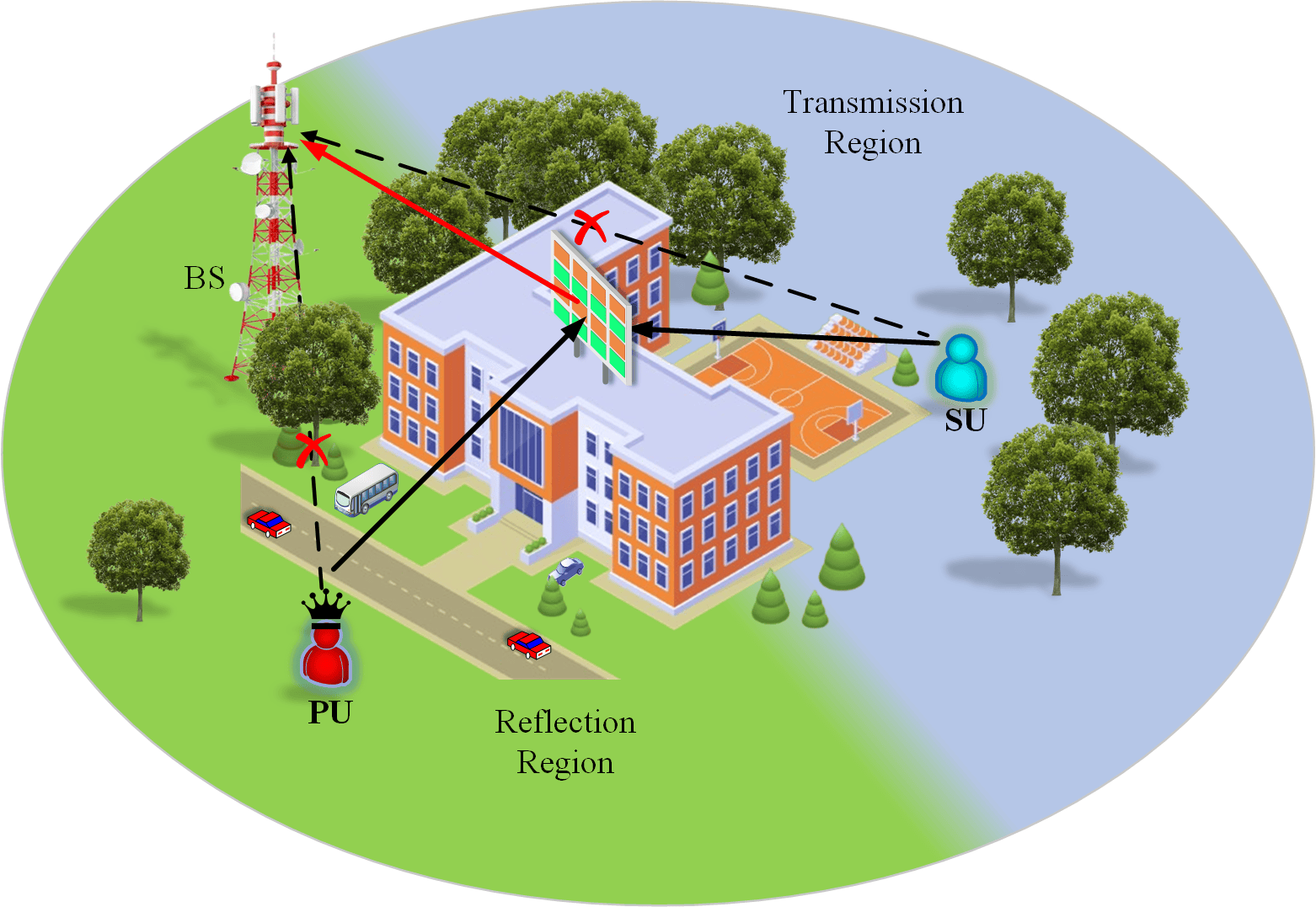}}
\caption{System overview.}
 \label{fig1}
 \end{figure}

To support this communication scenario, the PWE must ensure efficient signal transmission to the BS, while manipulating the transmissions appropriately to optimize network performance. In this direction, as shown in Fig.~\ref{fig1}, the PWE incorporates a STAR-RIS which is able to provide a $360^\circ$ field of view by simultaneously reflecting and transmitting electromagnetic waves. This capability ensures that users are served regardless of their positions relative to the STAR-RIS and overcomes the spatial constraints often encountered in dense urban deployments. In the considered scenario, the STAR-RIS performs an advanced electromagnetic functionality called \textit{CR functionality}, which enables dynamic wave manipulation to optimize spectral efficiency in CR scenarios. Specifically, it merges the concepts of element and power splitting by segmenting the STAR-RIS into two parts tailored to the PU's and SU's different requirements. This allows the proportion of power reflected or transmitted by each part to be controlled, thus modifying the channel gains of both the PU and SU and maximizing the efficiency of the CR-RSMA scheme. {\color{black} In more detail, the use of the STAR-RIS as an absorber is introduced and modeled by the parameters $\rho_{r}$ and $\rho_{t}$, which are power splitting factors indicating the fraction of power absorbed by the reflection and transmission elements, respectively. This mechanism is critical for operating the proposed STAR-RIS-enabled cognitive-based scheme because it allows for interference control through the STAR-RIS. Specifically, it can facilitate achieving the PU's target rate by reducing SU interference or increase the SU's data rate when the PU experiences a higher SINR than the target. Furthermore, this mechanism shifts the responsibility for interference control from the user, i.e., adapting $\alpha$, to the STAR-RIS, thus facilitating the fulfillment of the service level agreement.} 

Therefore, considering the network described in Fig.~\ref{fig1}, where a STAR-RIS with $N$ elements supports a CR-RSMA scenario through the CR functionality, the received signal at the BS can be expressed as
\begin{equation}
\small
\begin{split}
&y = \sqrt{P_p l_{pr} l_{br}} \rho_{r} \sum_{i=1}^{N_r} \left|\boldsymbol{h}_{i}^H\right|\big|\boldsymbol{h}_{pi}\big| e^{-j\phi_{i}^r} x_p \\
&\hspace{-0.16cm}+ \sqrt{P_s l_{sr} l_{br}} \rho_{t} \sum_{j=1}^{N_t} \left|\boldsymbol{h}_{j}^H\right|\big|\boldsymbol{h}_{sj}\big| e^{-j\phi_{j}^t} \left(\sqrt{\alpha} x_{s_1} + \sqrt{1-\alpha} x_{s_2}\right) \hspace{-0.08cm}+ z_n,
\end{split}
\label{eq1}
\end{equation}
where $P_p$ and $P_s$ denote the transmit powers of the PU and the SU, respectively,  $\alpha$ is the RS parameter. while $z_n$ represents the additive white Gaussian noise at the BS with zero mean and variance $\sigma_n^2$.
In addition, the channel from the PU to the $i$th reflecting element is represented by $\boldsymbol{h}_{pi}\in \mathbb{C}^{N\times 1}$ and the channel from the SU to the $j$th transmitting element by $\boldsymbol{h}_{sj}\in \mathbb{C}^{N\times 1}$, and the channels from the reflecting and transmitting elements to the BS are given by $\boldsymbol{h}_{i}^H\in \mathbb{C}^{1\times N}$ and $\boldsymbol{h}_{j}^H\in \mathbb{C}^{1\times N}$ respectively. Furthermore, the path losses between the PU and the STAR-RIS, the SU and the STAR-RIS, and the STAR-RIS and the BS are denoted by $l_{pr}$, $l_{sr}$, and $l_{br}$ respectively, with each expressed as $l_\epsilon = C_0\,d_\epsilon^{-\eta}$ where $C_0=\frac{\lambda^2}{16\pi^2}$ is the path loss at the reference distance $d_0=1\,\mathrm{m}$, $\lambda$ is the wavelength, and $\eta$ is the path loss exponent. The distance $d_\epsilon$, with $\epsilon\in \{pr, sr, br\}$, represents the separation between the corresponding nodes, where $d_{pr}$ is the distance between the PU and the STAR-RIS, $d_{sr}$ is the distance between the SU and the STAR-RIS, and $d_{br}$ is the distance between the BS and the STAR-RIS. In addition, the number of elements allocated for reflection is given by $N_r=\beta N$ and for transmission by $N_t=(1-\beta)N$, where $\beta \in [0,1] $ is the element allocation parameter. Finally, $\phi_{i}^r$ and $\phi_{j}^t$ are phase shift terms related to the channels established through the reflecting and the transmitting elements, respectively, and are given as $\phi_{i}^r = \vartheta_i^r + \arg(\boldsymbol{h}_{pi}) + \arg(\boldsymbol{h}_i^H)$ and $\phi_j^t = \vartheta_j^t + \arg(\boldsymbol{h}_{sj}) + \arg(\boldsymbol{h}_j^H)$, where $\vartheta_i^r$, $\vartheta_j^t$ denote the induced phase shifts from the reflecting and transmitting elements, and $\arg(\cdot)$ represents the argument of a complex number. 

To further strengthen the communication links between the PU and SU, the CR functionality integrates a beam-steering mechanism that configures each STAR-RIS element to cancel the phase distortions introduced by the wireless channels. Thus, by setting the induced phase shifts as $\vartheta_i^r = -\arg(\boldsymbol{h}_{pi}) - \arg(\boldsymbol{h}_i^H)$ for the reflecting elements and $\vartheta_j^t = -\arg(\boldsymbol{h}_{sj}) - \arg(\boldsymbol{h}_j^H)$ for the transmitting elements, the received signal $y$ can be rewritten as
\begin{equation}
\small
\begin{split}
&y = \sqrt{P_p l_{pr} l_{br}} \rho_{r} \sum_{i=1}^{N_r} \left|\boldsymbol{h}_{i}^H\right|\big|\boldsymbol{h}_{pi}\big| x_p \\
&\hspace{-0.1cm}+ \sqrt{P_s l_{sr} l_{br}} \rho_{t} \sum_{j=1}^{N_t} \left|\boldsymbol{h}_{j}^H\right|\big|\boldsymbol{h}_{sj}\big| \left(\sqrt{\alpha} x_{s_1} + \sqrt{1-\alpha} x_{s_2}\right)  + z_n .
\end{split}
\label{eq2}
\end{equation}
Thus, the PWE maximizes the efficiency of the CR-RSMA scheme by adjusting the element allocation parameter $\beta$, fine-tuning the RS parameter $\alpha$, and adjusting the individual power splitting values $\rho_{r}$ and $\rho_{t}$.

\subsection{Decoding Order \& Achievable Rate}
In conventional RSMA uplink systems, since the signals from both the PU and the SU arrive at the BS simultaneously, it is necessary to design an effective decoding strategy that not only mitigates interference, but also ensures that the QoS of the PU is preserved while maximizing the performance of the SU.  This is particularly critical in the CR-RSMA framework, where power splitting in RSMA allows the SU to adapt its transmission strategy for seamless coexistence with the PU. To further enhance this coexistence, CR-inspired successive interference cancellation (CR-SIC) is employed, which allows the receiver to progressively cancel interference during the decoding process, ensuring that the sequential decoding order effectively manages inter-user interference in a manner consistent with the goals of CR-RSMA \cite{YUE2023CR-RSMA}.

In this integrated framework, we adopt the decoding order $x_{s_1} \rightarrow x_p \rightarrow x_{s_2}$ as described in \cite{9755045}. In this sequence, the SU's first sub-message $x_{s_1}$ is decoded first, while $x_{s_2}$ is initially treated as interference to mitigate its negative effect on the decoding process. Next, the PU's message $x_p$ is decoded, with any remaining interference from $x_{s_2}$ effectively canceled through CR-SIC, and finally the SU's second sub-message $x_{s_2}$ is decoded. Throughout this process, the power allocation factor $\alpha$ is carefully chosen to regulate the interference encountered during the decoding of $x_p$, thus ensuring the performance of the SU and optimizing spectral efficiency. Accordingly, the received signal-to-interference-plus-noise ratio (SINR) expressions for the detection of $x_{s_1}$, $x_p$, and $x_{s_2}$ are given by
\begin{equation}
\small
\gamma_{s_1} = \frac{\alpha L_s g_s}{L_p g_p + (1-\alpha)L_s g_s + 1},
\label{su_snr1}
\end{equation}
\begin{equation}
\small
\gamma_p = \frac{L_p g_p}{(1-\alpha)L_s g_s + 1},
\label{pu_snr}
\end{equation}
\begin{equation}
\small
\gamma_{s_2} = (1-\alpha)L_s g_s,
\label{su_snr2}
\end{equation}
where the effective power factors $L_p$ and $L_s$ are defined as $L_p = P_p l_{pr}l_{br}$ and $L_s = P_s l_{sr}l_{br}$, and the effective channel gains are given by
\begin{equation}
\small
g_p = \rho_r^2 \left|\sum_{i=1}^{N_r} \left|\boldsymbol{h}_i^H\right|\left|\boldsymbol{h}_{pi}\right| \right|^2,
\label{gp}
\end{equation}
\begin{equation}
\small
g_s = \rho_t^2\left|\sum_{j=1}^{N_t} \left|\boldsymbol{h}_j^H\right|\left|\boldsymbol{h}_{sj}\right| \right|^2.
\label{gs}
\end{equation}

After establishing the decoding order and deriving the corresponding SINR expressions, the analysis naturally extends to evaluating the achievable rate performance. Since the achievable rate in a CR-RSMA system depends critically on effective interference management, ensuring that the QoS of the PU is maintained is of paramount importance. To achieve this, the effective SINR at the PU must exceed a target threshold, defined as $\hat{\gamma}_p = 2^{\hat{R}_p} - 1$, where $\hat{R}_p$ is the target rate of the PU. In addition, as indicated in \eqref{pu_snr}, the effective SINR of the PU is influenced by the prevailing channel conditions, which can be either robust or weak. Therefore, to classify the network state and determine the necessary adaptation strategy, we first evaluate whether the SINR requirement of the PU is satisfied when the power splitting factors are set as $\rho_r = \rho_t = 1$. This setting reflects the case where the STAR-RIS operates without power control, and serves to classify the underlying channel conditions as either favorable or adverse. Based on this classification, we identify distinct adaptation strategies, leading to the following case-based analysis derived from \eqref{pu_snr}:
\begin{itemize} 
\item $\frac{L_p g_p} {(1-\alpha) L_s g_s + 1} \geq \hat{\gamma}_p,~   \alpha \in [0,1]$:
When the PU encounters robust channel conditions, the resulting SINR can meet the target threshold.
\item $\frac{ L_p g_p} {(1-\alpha) L_s g_s + 1} \leq \hat{\gamma}_p,~\alpha \in [0,1]$: When the PU experiences weak channel conditions, the received SINR as show in \eqref{pu_snr} may fall below the target threshold. In this case, if decoding the PU's message $x_p$ while treating $x_{s_2}$ as interference does not meet the QoS requirement,
then considering that decoding $x_p$ without any interference from the SU. This situation can be further detailed as follows: 
    \begin{enumerate}[a)]
    \item If decoding $x_p$ without interference from the SU meets the target threshold, the PU service is maintained.
    \item If decoding $x_p$ without interference still does not meet the target threshold, the system may allocate all available resources to the SU, ensuring overall system performance and improved communication quality for the SU.
    \end{enumerate}
\end{itemize}
In summary, the conditions derived above establish a systematic basis for preserving the PU's QoS while enhancing the SU's performance. 
Below, we outline the corresponding rate expressions for these cases, thereby elucidating the capabilities of the CR functionality of STAR-RIS in RSMA networks: 

\textbf{Case I}: For the case where $L_p g_p \geq \hat{\gamma}_p\left((1-\alpha) L_s g_s+1\right)$ when $\rho_r = \rho_t = 1$, the PU's QoS constraint is satisfied, thus we can shift the focus to maximize the achievable rate of the SU. To accomplish this, we configure the power splitting factor $\rho_r$ to reduce the interference caused by the PU to the SU until the PU is brought to its SINR target. Consequently, the effective PU channel gain can be expressed as  
\begin{equation}
\small
\widetilde{g}_p = {\frac{\hat{\gamma}_p}{L_p}\Big((1-\alpha) L_s g_s + 1\Big)}.
\label{case1_gp}
\end{equation}
Therefore, by substituting \eqref{case1_gp} into \eqref{su_snr1}, the corresponding SNR for $x_{s_1}$ is formulated as
\begin{equation}
\small
\begin{aligned}
\widetilde{\gamma}_{s_1} &=  \frac{\alpha L_s g_s} {(1 + \hat{\gamma}_p)\left( (1-\alpha) L_s g_s + 1 \right)}.
\label{su_newsnr}
\end{aligned}
\end{equation}
Thus, considering both \eqref{su_snr1} and \eqref{su_snr2}, the total achievable rate of the SU in Case I is given as  
\begin{equation}
\small
\begin{split}
&R_s^{c1} = \log_2 \left( 1 +  \widetilde{\gamma}_{s_1} \right) + \log_2\Big(1 + (1-\alpha) L_s g_s \Big).
\label{rate_C1}
\end{split}
\end{equation}
\textbf{Case II}: If $\hat{\gamma}_p \leq  L_p g_p \leq \hat{\gamma}_p + \hat{\gamma}_p(1-\alpha) L_s g_s$ is satisfied for $\rho_r = \rho_t = 1$, then the PU does not meet its QoS requirement as its SINR remains below the target threshold $\hat{\gamma}_p$. 
However, by adjusting $\rho_t$, we can control the attenuation of the SU’s transmission and enable the PU to meet its target SINR. Under this configuration, the effective SU channel gain is given by
\begin{equation}
\small
\widetilde{g}_s = \frac{L_p g_p - \hat{\gamma}_p}{L_s(1-\alpha) \hat{\gamma}_p}.
\label{case2_gs}
\end{equation}

Thus, by plugging  \eqref{case2_gs} into \eqref{su_snr1} and \eqref{su_snr2} and after some algebraic manipulations, the sum of achievable rate for the SU in {Case II is formulated as
\begin{equation}
\small
R_s^{c2} = \log_2 \left( 1 +  \frac{\alpha\left(L_p g_p - \hat{\gamma}_p\right)}{(1-\alpha)L_p g_p(1+\hat{\gamma}_p)}\right) + \log_2 \left(\frac{L_p g_p}{\hat{\gamma}_p}\right).
 \label{rate_c2}
\end{equation}
\textbf{Case III}: When $L_p g_p \leq \hat{\gamma}_p$, the PU is unable to meet its QoS requirement, even under full power allocation and with the flexibility provided by CR-RSMA and CR-SIC. In such a scenario, where the PU remains in a persistent outage state, the PWE can adapt by reallocating all available resources to maximize the SU’s performance. To facilitate this, the CR functionality of the STAR-RIS is invoked to reconfigure the element allocation and power splitting strategy in favor of the SU. Specifically, the reflecting elements that were initially supporting the PU can be reassigned as transmitting elements, effectively converting the entire STAR-RIS surface into a transmission-focused mode that enhances the SU’s signal path. As a result, with the PU deactivated and only the SU occupying the channel, the SU’s achievable rate in Case III is given by
\begin{equation}
\small
R_s^{c3} = \log_2 \left( 1 + L_s g_s \right).
\label{rate_c3}
\end{equation}

It should be emphasized that Case III is a major difference of the proposed STAR-RIS-enabled CR-RSMA compared to the conventional CR-RSMA, where $\alpha$ is set equal to 1 when the PU is in outage. This leads to an improvement in the performance of the SU, since it is decoded without treating the PU as interference.

Thus, based on the identified cases, the achievable rate of the SU can be described as
\begin{align}\small
R_s = 
\begin{cases}
\vspace{0.2cm}
\log_2\left(1+ \zeta_1\,L_s g_s\right)\, & \textbf{Case I}, \\
\vspace{0.2cm}
\log_2\left(\left(1+ \zeta_2\right)\frac{L_p g_p}{\hat{\gamma}_p}-\zeta_2\right), & \textbf{Case II},  \\
\log_2 \left(1 +L_s g_s\right), & \textbf{Case III},  \\
\end{cases}
\label{su_rates}
\end{align}
where $\zeta_1 = (1-\alpha)+\frac{\alpha}{(1+\hat{\gamma}_p)}$ and $\zeta_2 = \frac{\alpha}{(1-\alpha)(1+\hat{\gamma}_p)}$.

\section{Rate Analysis}

In this section, we analyze the performance of the proposed STAR-RIS-assisted CR-RSMA system under realistic fading conditions. Specifically, considering Nakagami-$m$ fading, we derive closed-form expressions for the ergodic rate of the SU and the outage rate of the PU by leveraging statistical approximations of the effective channel gains.

\subsection{Statistical Channel Characterization}

To statistically characterize the fading environment in the considered STAR-RIS-assisted CR-RSMA system, the Nakagami-$m$ distribution is adopted, as it provides a flexible fading model that encompasses a wide range of propagation conditions, including both scattered and line-of-sight (LoS) dominated regimes. Specifically, the probability density function (PDF) of a Nakagami-$m$ distributed random variable (RV) is given as
\begin{equation}
\small
f_{n}(x) = \frac{2m^{m}} {\Gamma(m)\Omega^m} x^{2m-1} e^{-(\frac{m
}{\Omega}x^2)}, \quad x\geq0,
\end{equation}
where $m$ and $\Omega$ denote the shape and scale parameter of the distribution, respectively. In this direction, the user-to-RIS channel amplitudes $|\boldsymbol{h}_\varphi|$, with $\varphi \in \{p, s\}$, can be modeled as Nakagami-$m$ random variables with shape parameter $m$ and spread $\Omega$. Furthermore, given that the STAR-RIS is deployed at a high elevation with an unobstructed view toward the BS, as shown in Fig. 1, the RIS-to-BS channels are assumed to be deterministic, such that $|\boldsymbol{h}_i^H| = |\boldsymbol{h}_j^H| = 1$. Therefore, the effective channel gains \eqref{gp} and \eqref{gs} for the PU and SU can be rewritten as
\begin{equation}
\small
g_p = \rho_r^2 \left|\sum_{i=1}^{N_r} |\boldsymbol{h}_{pi}|\right|^2,
\end{equation}
and
\begin{equation}
\small
g_s = \rho_t^2 \left|\sum_{j=1}^{N_t} |\boldsymbol{h}_{sj}|\right|^2.
\end{equation}
These expressions involve the squared magnitude of coherent sums of Nakagami-$m$ distributed variables, which do not have closed-form distributions. However, since the number of reflecting elements is typically large, we approximate $g_p$ and $g_s$ using Gamma-distributed RVs obtained via moment-matching, which has been shown to yield tight and tractable approximations in such scenarios.
\textcolor{black}{To facilitate analysis,  a unified format of $g_p$ and $g_s$ can be expressed as a RV of $Z = \rho^2 \left| \sum\limits_{i=1}^{N} |\boldsymbol{h}_{\varphi i}| \right|^2$. }
\begin{proposition}
For sufficiently large $N$, the distribution of the RV $Z$, with $|\boldsymbol{h}_{\varphi i}|$ following Nakagami-$m$ distribution with shape and scale parameters $m$ and $\Omega$, respectively,  can be accurately approximated by a Gamma distribution, whose PDF is expressed as
\begin{equation}
\small
f_Z(z) = \frac{z^{k-1} e^{-z/\theta}}{\Gamma(k)\,\theta^k}, \quad z > 0,
\label{pdf}
\end{equation}
where $\Gamma(\cdot)$ is the Gamma function, and $k$ and $\theta$ denote the shape and scale parameters of the Gamma distribution, respectively, which are equal to
\begin{equation}
\small
k = \frac{I_1^2}{I_2 - I_1^2}, \quad \theta = \frac{\rho^2\left(I_2 - I_1^2\right)}{I_1},
\end{equation}
with $I_1$ and $I_2$ given as
\begin{equation}
\small
I_1 = N\left[\Omega + (N-1)\left(\frac{\Gamma\left(m+\frac{1}{2}\right)}{\Gamma(m)}\right)^2\frac{\Omega}{m}\right],
\label{moment1}
\end{equation}
and
\begin{equation}
\small
\begin{split}
 &I_2 = N\left[\frac{(m+1)\Omega^2}{m} +4(N-1)\left(\frac{\Gamma(m+\sfrac{3}{2})}{\Gamma(m)}\right)\left(\frac{\Omega}{m}\right)^{\sfrac{3}{2}} \right. \\
 &\left.
 + \, 3(N-1) \Omega^2 + 6(N-1)(N-2)\left(\frac{\Gamma(m+\sfrac{1}{2})}{\Gamma(m)}\right)^2 \frac{\Omega^2}{m} \right. \\
 & \left.
 \times(N-1)(N-2)(N-3)\left(\frac{\Gamma(m+\sfrac{1}{2})}{\Gamma(m)}\right)^4\left(\frac{\Omega}{m}\right)^2 \right].
\end{split}
\label{moment2}
\end{equation}
\end{proposition}
\begin{IEEEproof}
Considering that $N$ is large, by applying the moment-matching technique \cite{tyrharq,HQAM}, $\left| \sum\limits_{i=1}^{N} |\boldsymbol{h}_{\varphi i}| \right|^2$ can be approximated as a Gamma distributed RV with shape parameter $k'$ and scale parameter $\theta'$, which are equal to
\begin{equation}
\small
k' = \frac{I_1^2}{I_2 - I_1^2},
\end{equation}
and 
\begin{equation}
\small
\theta' = \frac{I_2 - I_1^2}{I_1}.
\end{equation}
Thus, as shown in \cite{HQAM}, the first and second moments of $\left| \sum\limits_{i=1}^{N} |\boldsymbol{h}_{\varphi i}| \right|^2$ are given by
\begin{equation}
\small
I_1 = N\left(\mathbb{E}\left[|h_{\varphi}|^2\right] + (N-1)\mathbb{E}^2\left[|h_{\varphi}|\right]\right),
\label{I1}
\end{equation}
and
\begin{equation}
\small
\begin{split}
&I_2 = N\Big( \mathbb{E}\left[|h_{\varphi}|^4\right] + 4(N-1)\mathbb{E}\left[|h_{\varphi}|^3\right] \mathbb{E}\left[|h_{\varphi}|\right] \\
& \quad + 3(N-1)\mathbb{E}^2\left[|h_{\varphi}|^2\right] + 6(N-1)(N-2)\mathbb{E}\left[|h_{\varphi}|^2\right] \mathbb{E}^2\left[|h_{\varphi}|\right] \\
& \quad + (N-1)(N-2)(N-3)\mathbb{E}^4\left[|h_{\varphi}|\right] \Big),
\end{split}
\label{I2}
\end{equation}
where $\mathbb{E}\left[\cdot\right]$ denotes expectation of an RV, and $\mathbb{E}\left[|h_{\varphi}|^n\right]$ is the $n$-th moment of a Nakagami-$m$ distributed RV and is equal to
\begin{equation}
\small
\mathbb{E}\left[|h_{\varphi}|^n\right] = \frac{\Gamma\left(m + \frac{n}{2}\right)}{\Gamma(m)}\left(\frac{\Omega}{m}\right)^{\frac{n}{2}}.
\label{moment_h}
\end{equation}
Therefore, by substituting \eqref{moment_h} into \eqref{I1} and \eqref{I2}, \eqref{moment1} and \eqref{moment2} can be derived. However, since the variable of interest is $
Z = \rho^2\left| \sum\limits_{i=1}^{N} |\boldsymbol{h}_{\varphi i}| \right|^2$, and noting that scaling a Gamma-distributed RV by a positive constant $\rho^2$ results in another Gamma-distributed RV with the same shape parameter and a scaled scale parameter, we obtain that $k = k'$, and $\theta = \rho^2 \theta'$, which concludes the proof.
\end{IEEEproof}
\subsection{SU Ergodic Rate}

Following the case-dependent expressions derived in \eqref{rate_C1}, \eqref{rate_c2}, and \eqref{rate_c3}, the ergodic performance of the SU is obtained by averaging the corresponding instantaneous rates over the joint distributions of the effective channel gains $g_s$ and $g_p$. By approximating $g_s$ and $g_p$ with the Gamma-distributed random variables $X$ and $Y$, respectively, where $X$ is characterized by a shape parameter $k_p=k$ and a scale parameter $\theta_p=\frac{\rho^2_{r}\left(I_2 - I_1^2\right)}{I_1}$, and $Y$ is characterized by a shape parameter $k_s=k$ and a scale parameter $\theta_s=\frac{\rho^2_{t}\left(I_2 - I_1^2\right)}{I_1}$, the total ergodic rate can be derived by integrating each case-specific rate expression over all possible fading realizations. Thus, the ergodic rate of the SU can be expressed as
\begin{equation} 
\small
     R_e = \text{ER}_1 + \text{ER}_2 + \text{ER}_3,
    \label{er}
\end{equation}
where $\text{ER}_1$, $\text{ER}_2$, and $\text{ER}_3$ denote the ergodic rates corresponding to Cases~I, II, and III, respectively, as determined by the underlying SINR conditions. In the following, each term is examined separately, and closed-form expressions are derived for the respective cases.
\begin{proposition}
The ergodic rate $\text{ER}_1$ of the SU in Case I can be tightly approximated in closed form as given in \eqref{er:c1_top}, where $\theta_{e}=\frac{I_2 - I_1^2}{I_1}$, $\psi={\hat{\gamma}_p(1-\alpha)L_s}/({L_p \theta_{e})}$, $\tau_1={\hat{\gamma}_p}/{(L_p \theta_{e})}$, and $\operatorname{Ei}(x) = -\int_{-x}^{\infty} \frac{e^{-t}}{t} \, dt$ denoting the exponential integral.
\end{proposition}
\begin{figure*}[t]
\hrule
\vspace{0.2cm}
\begin{equation}
\small
\begin{split}
\;&\text{ER}_1 \approx\ \frac{1}{\Gamma(k) \ln 2} \sum_{i_1=0}^{k-1} \sum_{n_1=0}^{i_1} \sum_{v_1=0}^{k+n_1-1} \binom{i_1}{n_1}\frac{(\theta_{e}\,\psi)^{n_1}\,\tau_1^{i_1 - n_1}\,e^{-\tau_1}}{i_1!\,({\psi\, \theta_{e} + 1})^{k + n_1}}  
\frac{(k + n_1 - 1)!}{(k + n_1 - v_1 - 1)!} \\[1ex]&\hspace{-0.1cm}
\times \left[\frac{(-1)^{k + n_1 - v_1 - 2}}{\left( \frac{\zeta_1\,L_s \theta_{e}}{\psi \theta_{e} + 1} \right)^{k + n_1 - v_1 - 1}} 
e^{\left( \frac{\psi\, \theta_{e} + 1}{\zeta_1 \,L_s\,\theta_{e}} \right)} \operatorname{Ei} \left( -\frac{\psi \,\theta_{e} + 1}{\zeta_1\,L_s \theta_{e}} \right) + \sum_{w_1=1}^{k + n_1 - v_1 - 1} (w_1 - 1)! \left( -\frac{\psi \,\theta_{e} + 1}{\zeta_1\,L_s \theta_{e}} \right)^{k + n_1 - v_1 - w_1 - 1} \right]
\end{split}
\label{er:c1_top}
\end{equation}
 \vspace{1ex}\hrule 
\end{figure*}
\begin{IEEEproof}
The ergodic rate of the SU in Case I based on \eqref{su_rates} is defined as
\begin{equation}
    \small
    \text{ER}_1 = \mathbb{E}\left[\log_2\left(1 + \zeta_1\,L_s g_s\right)\right].
    \label{eq_initialER1}
\end{equation}
As established during the case identification process, the SINR condition characterizing Case I involves the PU and SU channel gains, both evaluated under the assumption of $\rho_r = \rho_t = 1$. In this direction, by considering the PDFs of $X$ and $Y$ corresponding to the PU and SU channel gains for $\rho_r = \rho_t = 1$, \eqref{eq_initialER1} can be expressed as
\begin{equation}
\small
\begin{split}
&\text{ER}_1 \approx \frac{1}{\Gamma(k)^2\, \theta_{e}^{2k}} \\& \times \int_0^{\infty} \int_{\frac{\hat{\gamma}_p}{L_p} \left((1 - \alpha) L_s y+1\right)}^{\infty} \hspace{-0.6cm}\log_2(1 + \zeta_1 L_s y)\, (xy)^{k-1}e^{-\frac{x+y}{\theta_{e}}}dx\, dy,
\end{split}
\label{er1_1}
\end{equation}
where $\theta_{e}=\frac{\left(I_2 - I_1^2\right)}{I_1}$. Additionally, by utilizing the definition of upper incomplete Gamma function $\Gamma(k, z)$~\cite[Eq.~(8.350.2)]{Ryzhik2014}, converting $\log_2(\cdot)$ into natural logarithm form, and setting $\tau_1=\frac{\hat{\gamma}_p}{L_p \theta_{e}}$, and $\psi=\frac{\hat{\gamma}_p(1-\alpha)\,L_s}{L_p\theta_{e}}$, \eqref{er1_1} can be reformulated as
\begin{equation}
\small
\text{ER}_1 \! \approx \! \frac{1}{\ln(2)\,\Gamma(k)^2\theta^k_{e}}\hspace{-0.14cm}\int_0^{\infty}\hspace{-0.19cm} \Gamma(k,\tau_1+\psi y)\,\ln(1+\zeta_1 L_s y)\,y^{k-1}\, \hspace{-0.1cm}e^{-y/\theta_{e}}\,dy.
\label{A1_integral}
\end{equation}
Moreover, by utilizing the series expansion of upper incomplete Gamma function ~\cite[Eq.~(8.352.4)]{Ryzhik2014}, \eqref{A1_integral} can be rewritten as
\begin{equation}
\small
\begin{split}
&\text{ER}_1 \approx \frac{e^{-\tau_1}}{\ln(2)\, \Gamma(k)^2} \sum_{i_1 = 0}^{k-1} \sum_{n_1 = 0}^{i_1} \frac{\tau_1^{i_1 - n_1} \binom{i_1}{n_1} (\theta_{e} \psi)^{n_1}(k - 1)!}{i_1! \left(\theta_{e} \psi + 1\right)^{k + n_1}}\\&
\times \int_0^{\infty} e^{-y} y^{k + n_1 - 1} \ln\left(1 + \frac{\zeta_1 L_s \theta_{e}}{\theta_{e} \psi + 1} y \right)\, dy.
\end{split}
\end{equation}
Finally, by utilizing \cite[Eq. (4.337.5)]{Ryzhik2014}, and after some algebraic manipulations, \eqref{er:c1_top} can be derived, which concludes the proof.
\end{IEEEproof}

Next, a closed-form expression of the SU’s ergodic rate corresponding to Case II is derived in the following proposition.
\begin{proposition}
The ergodic rate $\text{ER}_2$ of the SU in Case II can be tightly approximated as in \eqref{er:c2}, where $\tau_1={\hat{\gamma}_p}/{(L_p\theta_{e})}$, $\tau_2 = \tau_1/(\zeta_2+1)$ and $\gamma(\cdot,\cdot)$ represents the lower incomplete gamma function.
\end{proposition}
\begin{figure*}[t]
\begin{equation}
\small
\begin{split}
&\text{ER}_2\approx\frac{e^{-\frac{\zeta_2 \,\tau_1}{\zeta_2+1}}}{\Gamma (k)\ln 2} \sum_{i_2=0}^{k-1} \binom{k-1}{i_2} (\zeta_2\, \tau_2)^{i_2} \left\{\Gamma (\kappa) \sum_{q_2 = 0}^{k-1} \frac{1}{(k-q_2-1)!} \right. \\[1ex]&\left. \times\left[\sum_{t_2 = 1}^{k-q_2-1} (t_2-1)! \left(-\frac{\tau_2}{\theta_{e} \psi}\right)^{k-q_2-t_2-1} \hspace{-0.15cm}+ \frac{(-1)^{k-q_2-2} e^{\frac{\tau_2}{\theta_{e} \psi}} \operatorname{Ei}\left(-\frac{\tau_2}{\theta_{e} \psi}\right)}{\left(\frac{\theta_{e} \psi }{\tau_2}\right)^{k-q_2-1}}\right] - \sum_{a_2=0}^{\kappa-1} \sum_{n_2=0}^{a_2}\binom{a_2}{n_2} \frac{e^{-\tau_2} (\kappa-1)!\, \tau_2^{a_2-n_2} (\theta_{e} \psi )^{n_2}}{a_2!\, \Gamma(k) (\theta_{e} \psi+1)^{k+n_2}} \right. \\[1ex]&\left.
\times \left[\ln(\tau_2)\, \Gamma(k+n_2)+\sum_{e_2=0}^{k+n_2-1} \frac{(k+n_2-1)!}{(k+n_2-e_2-1)!} \left(\sum_{r_2=1}^{k+n_2-e_2-1} (r_2-1)! \left(-\frac{\tau_2 (\theta_{e} \psi +1)}{\theta_{e} \psi }\right)^{k+n_2-e_2-r_2-1} +e^{\frac{\tau_2 (\theta_{e} \psi +1)}{\theta_{e} \psi }}\right. \right. \right. \\[1ex]&\left. \left. \left. \times \frac{(-1)^{k+n_2-e_2-2}}{\left(\frac{\theta_{e} \psi }{\tau_2 (\theta_{e} \psi +1)}\right)^{k+n_2-e_2-1}} \hspace{0.15cm} \operatorname{Ei}\left(-\frac{\tau_2(\theta_{e} \psi +1)}{\theta_{e} \psi}\right)\right)\right]-\sum_{b_2=0}^{\kappa-1} \sum_{c_2=0}^{b_2}\binom{b_2}{c_2}\,\ln \left(\frac{1}{\tau_2}\right) \times \frac{e^{-\tau_2}  (\kappa-1)! \, \tau_2^{b-c} \, (\theta_{e} \psi )^{c_2} \, (k)_{c_2}}{{b_2}! \,(\theta_{e} \psi +1)^{k+c_2}} \right. \\[1ex]&\left.
-\sum_{j_2=0}^{\infty} \sum_{f_2=0}^{\kappa+j_2}\binom{\kappa+j_2}{f_2} \frac{(-1)^{j_2} \tau_2^{\kappa+j_2-f_2} \,(\theta_{e} \psi )^{f_2}\,(k)_{f_2}}{{j_2}! \, ({\kappa+j})^2}  +\frac{\tau_2^{\kappa}}{\kappa^2} \, {}_2F_2(\kappa,\kappa;\kappa+1,\kappa+1;\tau_2)
\right\}
  \end{split}
  \label{er:c2}
  \end{equation}
  \vspace{1ex}\hrule 
\end{figure*}

\begin{IEEEproof}
The proof is provided in Appendix \ref{case2append}.
\end{IEEEproof}

Finally, in the following proposition, we provide a closed-form expression for the ergodic rate of the SU in Case III.
\begin{proposition}\label{prop:case3}
The ergodic rate $\text{ER}_3$ of the SU in Case III can be tightly approximated as
\begin{equation}
\small
\begin{split}
&\text{ER}_3 \approx \frac{\gamma(k,\tau_1)}{\Gamma(k)^2 \ln 2} \sum_{i_3 = 0}^{k - 1} \frac{(k - 1)!}{(k - i_3 - 1)!}  \times \Bigg[\frac{(-1)^{k - i_3 - 2}}{(L_s\theta_{e})^{k - i_3 - 1}} e^{1/(L_s\theta_{e})} \\& \times \operatorname{Ei}\left(-\frac{1}{L_s\theta_{e}}\right)+ \sum_{j_3 = 1}^{k - i_3 - 1} (j_3 - 1)! \left(-\frac{1}{L_s\theta_{e}}\right)^{k - i_3 - j_3 - 1} \Bigg].
\end{split}
\label{er:c3}
\end{equation}
\end{proposition}
\begin{IEEEproof}
Similarly with the previous cases, the ergodic rate of the SU in Case III according to \eqref{su_rates} can be expressed as
\begin{equation}
\small
\text{ER}_3 = \int_0^\infty \int_0^{\hat{\gamma}_p/L_p} \log_2(1 + L_s y)\frac{(x y)^{k-1}\,e^{-\frac{x+y}{\theta_{e}}}}{\Gamma(k)^2\, \theta_{e}^{2k}} \,dx\, dy.
\label{er3}
\end{equation}
By utilizing the integral form of the lower incomplete gamma function $\gamma(k, \tau_1)$, \eqref{er3} can be rewritten as
\begin{equation}
\small
\text{ER}_3 \approx \frac{\gamma(k, \tau_1)}{\Gamma(k)^2\,\ln 2 } \int_0^\infty \ln(1 + L_s\theta_e\, y) \,y^{k-1} e^{-y}\,dy.
\end{equation}
Finally, by utilizing \cite[Eq. (4.337.5)]{Ryzhik2014} and after some algebraic manipulations, \eqref{er:c3} can be obtained, which concludes the proof.
\end{IEEEproof}
\begin{remark}
The ergodic rate expression in \eqref{er}, along with the closed-form results provided in Propositions 2--4, is derived for a general RS parameter $\alpha \in [0,1]$, providing a flexible analytical framework for STAR-RIS-assisted CR-RSMA systems. For scenarios that require optimizing the SU's ergodic rate, $\alpha$ can be appropriately selected based on the decoding strategy, as discussed in~\cite{YUE2023CR-RSMA}.
\end{remark}

\subsection{PU Outage Rate}
In CR systems, the PU is typically subject to strict QoS requirements, often expressed as a target SNR threshold that must be maintained to ensure reliable communication.
In this context, the outage rate is a key performance metric that reflects the PU's ability to meet its QoS constraint in the presence of interference and dynamic fading conditions. 
\textcolor{black}{
Specifically, the outage rate of PU is defined as 
\begin{equation}
\small
R_o  = \hat{R}_p (1-P_{out}) ,
\label{or}
\end{equation}
where $\hat{R}_p=\log_2(1+\hat{\gamma}_p)$ denotes the PU target rate and $P_{out}$ is the outage probability of the PU, which can be expressed as
\begin{equation}
\small
P_{\text{out}} = \text{Pr}\left(\gamma_p \leq \hat{\gamma}_p\right).
\label{op} 
\end{equation}
Herein, \eqref{op}
denotes the probability that the instantaneous SNR at the PU falls below a predefined target value, indicating a transmission failure and enabling accurate quantification of the system's behavior under varying degrees of interference and channel degradation.}

\begin{proposition}
The outage rate $R_o$ of the PU can be approximated in closed form as
\begin{equation}
\small
\begin{split}
R_o \approx \frac{\ln(1+\hat{\gamma}_p)}{\ln 2}\sum_{i_o=0}^{k-1}\sum_{j_o=0}^{i_o}\binom{i_o}{j_o}\frac{\tau_1^{i_o-j_o} (\psi\,\theta_e)^{j_o} e^{-\tau_1}}{(\theta_e\, \psi+1)^{k+j_o}\,{i_o}!} (k)_{j_o}
\end{split}
\label{or:cf}
\end{equation}
where $(k)_{j_o}=\frac{\Gamma(k+j_o)}{\Gamma(k)}$ is the Pochhammer symbol.
\end{proposition}
\begin{IEEEproof}
By utilizing \eqref{pu_snr}, \eqref{op} can be rewritten as
\begin{equation}
\small
    P_{\text{out}} = \text{Pr}\left(g_p\leq\frac{\hat{\gamma}_p}{L_p}\bigl((1-\alpha)L_s g_s + 1\bigr)\right).
    \label{pout}
\end{equation}
Moreover, by expressing \eqref{pout} in integral form, we obtain
\begin{equation}
\small
P_{\text{out}} \approx \int_0^\infty \int_0^{\frac{\hat{\gamma}p}{L_p}\bigl((1 - \alpha)L_sy+1\bigr)} \frac{(x y)^{k-1}\,e^{-\frac{x+y}{\theta_{e}}}}{\Gamma(k)^2\theta_{e}^{2k}} \,dx\, dy,
\label{op:int}
\end{equation}
\color{black}
and, by invoking the lower incomplete Gamma function definition ~\cite[Eq.~(8.350.2)]{Ryzhik2014}, \eqref{op:int} can be rewritten as
\begin{equation} 
\small
P_{\text{out}} \approx \frac{1}{\Gamma(k)^2\theta_e^k} \int_0^\infty \gamma\left(k, \tau_1 + \psi y \right) y^{k-1}e^{-y/\theta_e} \, dy.
\label{prof:out1} 
\end{equation}
Additionally, by applying the identity $\gamma(k, z) = \Gamma(k) - \Gamma(k, z)$ and utilizing the series representation of the upper incomplete Gamma function ~\cite[Eq.~(8.352.4)]{Ryzhik2014}, \eqref{prof:out1} is reformulated as
\begin{equation} 
\small 
\begin{split}
&P_{\text{out}} \approx1-\sum _{i_o=0}^{k-1} \frac{e^{-\tau }}{\Gamma(k) {i_o}!} \int_0^{\infty} (\tau_1 +\theta_e \psi y)^{i_o} y^{k-1} e^{-(\theta_e \psi +1)y} dy. 
\end{split}
\label{prof:out2} 
\end{equation}
Finally, by utilizing the definitions of Gamma function and of the Pochhammer symbol, after some algebraic manipulations, \eqref{or:cf} is obtained, thus concluding the proof.
\end{IEEEproof}

\section{Simulation Results}
In this section, we evaluate the performance of the CR functionality in the examined scenario, as well as the accuracy and validity of the derived expressions. Specifically, we consider a total distance $d$ between the PU and the SU, and define the distance parameter $\omega \in [0, 1]$ to control user positioning relative to the STAR-RIS, where the distance from the PU to the STAR-RIS is set as $d_1 = \omega d$, while the SU is located at $d_2 = (1 - \omega) d$, respectively. In addition, Table~\ref{table} includes the selected values of the parameters considered in the examined scenario, while unless otherwise stated, the RS parameter $\alpha$ is set to $0.7$, the element allocation parameter $\beta$ is set equal to $0.3$, the PU target threshold $\hat{\gamma}_p$ is set equal to $3$, and the distance parameter $\omega$ is set equal to $0.2$. Finally, to evaluate the accuracy of the derived expressions, we perform Monte Carlo simulations with $10^6$ realizations to verify the accuracy of the derived analytical results. To this end, the simulation results are illustrated as marks and the theoretical results are illustrated as solid lines.

\color{black}
\begin{table}
    \centering
    \caption{Simulation Result Parameters}
    \begin{tabular}{p{4.5cm} p{1.5cm} p{1cm}}
           \hline
           \textbf{Parameter} & \textbf{Notation}
           & \textbf{Value} \\
           \hline\hline
           Pathloss exponent & $\eta$ & $2$\\ 
           Number of STAR-RIS elements & $N$ & $400$\\ 
            Total distance & $d$& $30\,\mathrm{m}$ \\
            Noise power & $\sigma$ & $-60\,\mathrm{dB}$\\ 
            Shape parameter &  $m$& $3$ \\ 
            scale parameter & $\Omega$& $1$ \\ 
            Reference distance & $d_0$ & $1\,\mathrm{m}$ \\ 
            Pathloss @ reference distance & $C_o$ & $10\,\mathrm{mW}$\\ 
            Transmit power for the PU & $P_p$ & $10\,\mathrm{dBm}$\\ 
            Transmit power for the SU & $P_s$ & $30\,\mathrm{dBm}$ \\ \hline
        \end{tabular}
         \label{table}
        \end{table}

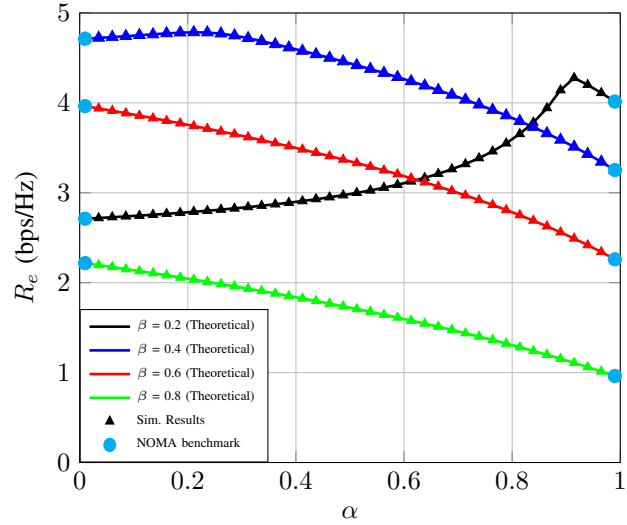
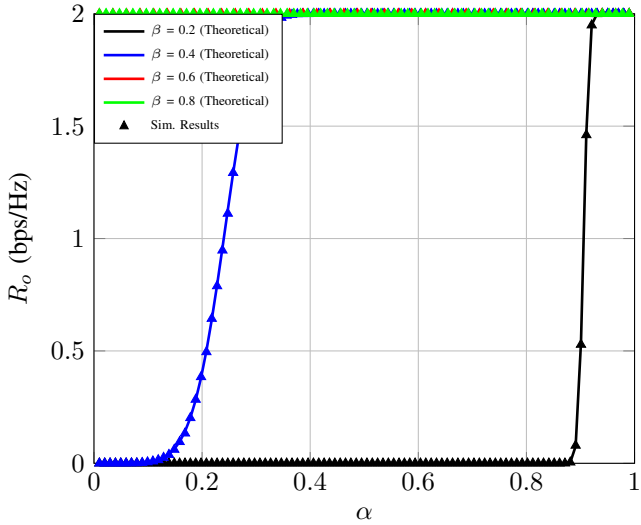
\begin{figure}[h!]
\centering
\begin{subfigure}{\linewidth}
\centering
\begin{tikzpicture}
\begin{axis}[width=0.99\linewidth,
xlabel = {$\alpha$},
ylabel = {$R_e$ (bps/Hz)},
xmin = 0, xmax = 1,
ymin = 0, ymax = 5,
xtick = {0,0.2,...,1},
ytick = {0,1,...,5},
grid = major, legend image post style={xscale=0.9, every mark/.append style={solid}},
legend cell align = {left},
legend style={at={(0,0)}, anchor=south west, font = \tiny} ]
\addplot[
black,
no marks,
line width = 1pt,
solid]
table{Figures/alpha/ER_fix_omega/beta0.2_cf.txt}; \addlegendentry{$\beta$ = 0.2 (Theoretical)}
\addplot[
blue,
no marks,
line width = 1pt,
solid]
table{Figures/alpha/ER_fix_omega/beta0.4_cf.txt}; \addlegendentry{$\beta$ = 0.4 (Theoretical)}
\addplot[
red,
no marks,
line width = 1pt,
solid]
table{Figures/alpha/ER_fix_omega/beta0.6_cf.txt}; \addlegendentry{$\beta$ = 0.6 (Theoretical) }
\addplot[
green,
no marks,
line width = 1pt,
solid]
table{Figures/alpha/ER_fix_omega/beta0.8_cf.txt}; \addlegendentry{$\beta$ = 0.8 (Theoretical) }

\addplot[
black,
only marks,
mark=triangle*,
mark size = 2,]
table{Figures/alpha/ER_fix_omega/beta0.2_mc.txt}; 
\addlegendentry{Sim. Results}

\addplot[
cyan,
only marks,
mark=*,
mark size = 2.5,]
table{Figures/alpha/noma_alpha_0/noma_0.2.txt}; \addlegendentry{NOMA benchmark}

\addplot[
blue,
only marks,
mark=triangle*,
mark size = 2.5,]
table{Figures/alpha/ER_fix_omega/beta0.4_mc.txt}; 
\addplot[
red,
only marks,
mark=triangle*,
mark size = 2,]
table{Figures/alpha/ER_fix_omega/beta0.6_mc.txt}; 
\addplot[
green,
only marks,
mark=triangle*,
mark size = 2,]
table{Figures/alpha/ER_fix_omega/beta0.8_mc.txt};

\addplot[
cyan,
only marks,
mark=*,
mark size = 2.5,]
table{Figures/alpha/noma_alpha_0/noma_0.4.txt};
\addplot[
cyan,
only marks,
mark=*,
mark size = 2.5,]
table{Figures/alpha/noma_alpha_0/noma_0.6.txt};
\addplot[
cyan,
only marks,
mark=*,
mark size = 2.5,]
table{Figures/alpha/noma_alpha_0/noma_0.8.txt};
\addplot[
cyan,
only marks,
mark=*,
mark size = 2.5,]
table{Figures/alpha/noma_alpha_1/noma_0.2.txt}; 

\addplot[
cyan,
only marks,
mark=*,
mark size = 2.5,]
table{Figures/alpha/noma_alpha_1/noma_0.4.txt};
\addplot[
cyan,
only marks,
mark=*,
mark size = 2.5,]
table{Figures/alpha/noma_alpha_1/noma_0.6.txt};
\addplot[
cyan,
only marks,
mark=*,
mark size = 2.5,]
table{Figures/alpha/noma_alpha_1/noma_0.8.txt};
\end{axis}
\end{tikzpicture}
\subcaption{Ergodic rate of the SU.}
\end{subfigure}
\vspace{0.5cm}
\begin{subfigure}{\linewidth}
\centering
\begin{tikzpicture}
\begin{axis}[width=0.99\linewidth,
xlabel = {$\alpha$},
ylabel = {$R_o$ (bps/Hz)},
xmin = 0, xmax = 1,
ymin = 0, ymax = 2,
xtick = {0,0.2,...,1},
grid = major, legend image post style={xscale=0.9, every mark/.append style={solid}},
legend cell align = {left},
legend style={at={(0,1)}, anchor=north west, font = \tiny} ]
\addplot[
black,
no marks,
line width = 1pt,
solid]
table{Figures/alpha/OR_fix_omega/beta0.2_cf.txt}; \addlegendentry{$\beta$ = 0.2 (Theoretical)}

\addplot[
blue,
no marks,
line width = 1pt,
solid]
table{Figures/alpha/OR_fix_omega/beta0.4_cf.txt}; \addlegendentry{$\beta$ = 0.4 (Theoretical)}
\addplot[
red,
no marks,
line width = 1pt,
solid]
table{Figures/alpha/OR_fix_omega/beta0.6_cf.txt}; \addlegendentry{$\beta$ = 0.6 (Theoretical)}

\addplot[
green,
no marks,
line width = 1pt,
solid]
table{Figures/alpha/OR_fix_omega/beta0.8_cf.txt}; \addlegendentry{$\beta$ = 0.8 (Theoretical)}

\addplot[
black,
only marks,
mark=triangle*,
mark size = 2,]
table{Figures/alpha/OR_fix_omega/beta0.2_mc.txt}; \addlegendentry{Sim. Results}
\addplot[
blue,
only marks,
mark=triangle*,
mark size = 2,]
table{Figures/alpha/OR_fix_omega/beta0.4_mc.txt}; 
\addplot[
red,
only marks,
mark=triangle*,
mark size = 2,]
table{Figures/alpha/OR_fix_omega/beta0.6_mc.txt};
\addplot[
green,
only marks,
mark=triangle*,
mark size = 2,]
table{Figures/alpha/OR_fix_omega/beta0.8_mc.txt};
\end{axis}
\end{tikzpicture}
\caption{Outage rate of the PU.}
\end{subfigure}
    \caption{Ergodic rate and outage rate versus $\alpha$ for various $\beta$ values with $\omega = 0.2$.}
    \label{fig2}
\end{figure}

Figs. \ref{fig2}a and \ref{fig2}b illustrate how the RS parameter $\alpha$ affects the performance of the proposed STAR-RIS-assisted CR-RSMA system, evaluated for various values of the element allocation parameter $\beta$, in the case where the STAR-RIS is positioned near the PU. Specifically, Fig. \ref{fig2}a depicts the ergodic rate achieved by the SU, while Fig. \ref{fig2}b presents the corresponding outage rate of the PU. As observed, the derived theoretical expressions coincide with the simulation results, confirming their accuracy in capturing the behavior of the CR functionality across different configurations. When $\beta$ is set to $0.2$, with most elements assigned to transmission, the SU exhibits performance that closely aligns with that of a NOMA-based system, and achieves its highest ergodic rate at larger $\alpha$ values due to better signal delivery to its private message. However, the limited number of reflecting elements prevents the PU from satisfying its QoS constraint throughout the entire $\alpha$ range. As $\alpha$ increases, the interference from the SU’s second sub-message decreases, allowing the PU to briefly meet its target only when the SU allocates nearly all its power to the common message. When $\beta$ increases to $0.4$, more elements are dedicated to reflection, enhancing the PU’s channel and enabling it to meet its outage threshold when $\alpha$ exceeds approximately $0.2$, while still maintaining a high ergodic rate for the SU, which remains comparable to NOMA performance. This demonstrates how the CR functionality of the STAR-RIS adaptively reconfigures the surface to serve both users by shaping the propagation environment in response to interference and channel conditions. Finally, for $\beta = 0.6$ and $0.8$, the number of reflecting elements becomes sufficient to support the PU across all values of $\alpha$, resulting in a consistently satisfied outage probability requirement. In this regime, the PU becomes effectively decoupled from variations in $\alpha$, granting the SU greater flexibility to optimize its RS parameter. However, the reduced number of transmitting elements leads to a gradual degradation in SU performance, which falls below that of NOMA for all $\alpha$ values. Thus, Figs. \ref{fig2}a and \ref{fig2}b underscore the importance of jointly tuning $\alpha$ and $\beta$, as user performance is directly influenced by the dynamic balance between transmission and reflection achieved through the CR functionality of the STAR-RIS.

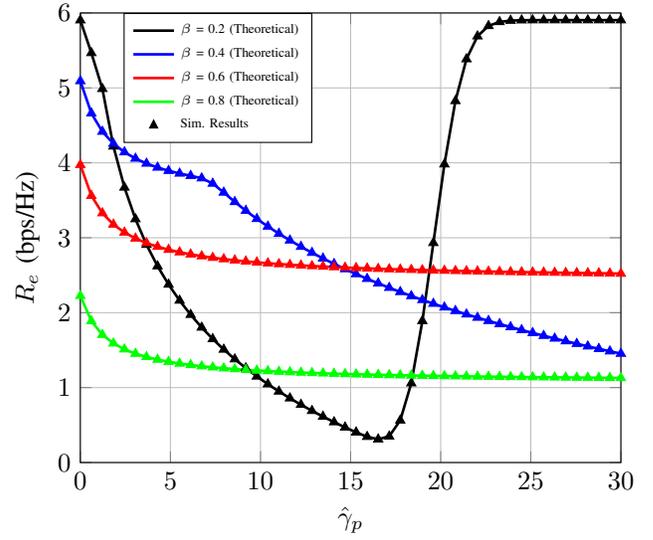
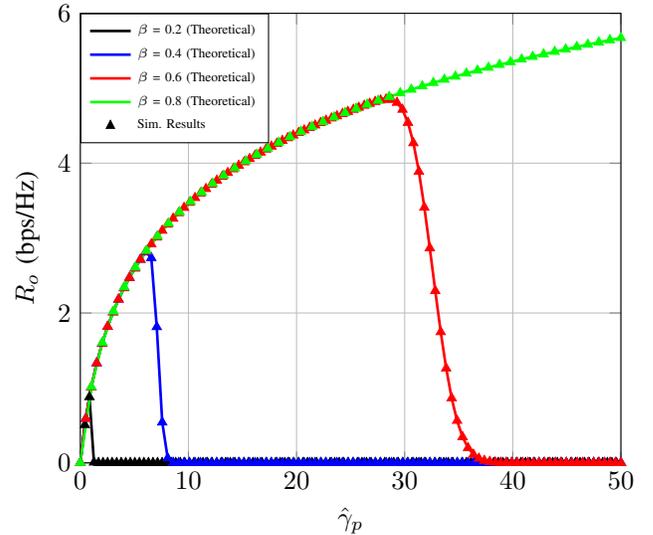
\begin{figure}[h!]
\centering
\begin{subfigure}{\linewidth}
\centering
\begin{tikzpicture}
\begin{axis}[width=0.99\linewidth,
xlabel = {$\hat{\gamma}_p$},
ylabel = {$R_e$ (bps/Hz)},
xmin = 0, xmax = 30,
ymin = 0, ymax = 6,
xtick = {0,5,...,30},
ytick = {0,1,...,6},
grid = major, legend image post style={xscale=0.9, every mark/.append style={solid}},
legend cell align = {left},
legend style={at={(0.08,1)}, anchor=north west, font = \tiny} ]
\addplot[
black,
no marks,
line width = 1pt,
solid]
table{Figures/gamma/ER_fix_omega/beta0.2_cf.txt}; \addlegendentry{$\beta$ = 0.2 (Theoretical)}
\addplot[
blue,
no marks,
line width = 1pt,
solid]
table{Figures/gamma/ER_fix_omega/beta0.4_cf.txt}; \addlegendentry{$\beta$ = 0.4 (Theoretical)}
\addplot[
red,
no marks,
line width = 1pt,
solid]
table{Figures/gamma/ER_fix_omega/beta0.6_cf.txt}; \addlegendentry{$\beta$ = 0.6 (Theoretical) }
\addplot[
green,
no marks,
line width = 1pt,
solid]
table{Figures/gamma/ER_fix_omega/beta0.8_cf.txt}; \addlegendentry{$\beta$ = 0.8 (Theoretical) }
\addplot[
black,
only marks,
mark=triangle*,
mark size = 2,]
table{Figures/gamma/ER_fix_omega/beta0.2_mc.txt}; 
\addlegendentry{Sim. Results}
\addplot[
blue,
only marks,
mark=triangle*,
mark size = 2,]
table{Figures/gamma/ER_fix_omega/beta0.4_mc.txt}; 
\addplot[
red,
only marks,
mark=triangle*,
mark size = 2,]
table{Figures/gamma/ER_fix_omega/beta0.6_mc.txt}; 
\addplot[
green,
only marks,
mark=triangle*,
mark size = 2,]
table{Figures/gamma/ER_fix_omega/beta0.8_mc.txt}; 
\end{axis}
\end{tikzpicture}
\subcaption{Ergodic rate of the SU.}
\end{subfigure}
\vspace{0.5cm}
\begin{subfigure}{\linewidth}
\centering
\begin{tikzpicture}
\begin{axis}[width=0.99\linewidth,
xlabel = {$\hat{\gamma}_p$},
ylabel = {$R_o$ (bps/Hz)},
xmin = 0, xmax = 50,
ymin = 0, ymax = 6,
xtick = {0,10,...,50},
grid = major, legend image post style={xscale=0.9, every mark/.append style={solid}},
legend cell align = {left},
legend style={at={(0,1)}, anchor=north west, font = \tiny} ]
\addplot[
black,
no marks,
line width = 1pt,
solid]
table{Figures/gamma/OR_fix_omega/beta0.2_cf.txt}; \addlegendentry{$\beta$ = 0.2 (Theoretical)}
\addplot[
blue,
no marks,
line width = 1pt,
solid]
table{Figures/gamma/OR_fix_omega/beta0.4_cf.txt}; \addlegendentry{$\beta$ = 0.4 (Theoretical)}
\addplot[
red,
no marks,
line width = 1pt,
solid]
table{Figures/gamma/OR_fix_omega/beta0.6_cf.txt}; \addlegendentry{$\beta$ = 0.6 (Theoretical)}

\addplot[
green,
no marks,
line width = 1pt,
solid]
table{Figures/gamma/OR_fix_omega/beta0.8_cf.txt}; \addlegendentry{$\beta$ = 0.8 (Theoretical)}
\addplot[
black,
only marks,
mark=triangle*,
mark size = 2,]
table{Figures/gamma/OR_fix_omega/beta0.2_mc.txt}; \addlegendentry{Sim. Results}
\addplot[
blue,
only marks,
mark=triangle*,
mark size = 2,]
table{Figures/gamma/OR_fix_omega/beta0.4_mc.txt};
\addplot[
red,
only marks,
mark=triangle*,
mark size = 2,]
table{Figures/gamma/OR_fix_omega/beta0.6_mc.txt};
\addplot[
green,
only marks,
mark=triangle*,
mark size = 2,]
table{Figures/gamma/OR_fix_omega/beta0.8_mc.txt};
\end{axis}
\end{tikzpicture}
\caption{Outage rate of the PU.}
\end{subfigure}
    \caption{Ergodic rate and outage rate versus $\hat{\gamma}_p$ for various $\beta$ values with $\omega = 0.2$.}
    \label{fig4}
\end{figure}

Figs. \ref{fig4}a and \ref{fig4}b illustrate the impact of the PU’s target threshold $\hat{\gamma}_p$ on the performance of the proposed STAR-RIS-assisted CR-RSMA system, evaluated for different values of the element allocation parameter $\beta$, under a deployment where the STAR-RIS is positioned near the PU. Specifically, Fig. \ref{fig4}a presents the ergodic rate of the SU, while Fig. \ref{fig4}b shows the corresponding outage rate of the PU. As can be seen, when $\beta$ is set to $0.2$, with most elements allocated to transmission, the SU initially benefits from a high ergodic rate, while the PU operates within its required outage constraint. As $\hat{\gamma}_p$ increases, the SINR needed by the PU becomes more stringent, progressively limiting the system’s ability to maintain the PU’s QoS and leading to a sharp rise in the outage probability. Once this threshold exceeds the system's support capability, the STAR-RIS dynamically reallocates all available resources to the SU, resulting in an abrupt increase in its ergodic rate. This shift, reflected in both the analytical and simulated curves, showcases how the STAR-RIS, through its CR functionality, adapts its configuration to maximize SU performance when PU protection becomes infeasible. In addition, when $\beta$ is increased, more elements are reserved for reflection, enabling the PU to meet higher threshold demands while maintaining low outage probability over a broader range. However, this improved PU support comes at the cost of a reduced SU rate, which remains nearly flat across the entire $\hat{\gamma}_p$ range for large $\beta$ values, as seen in Fig. \ref{fig4}a. Therefore, Figs. \ref{fig4}a and \ref{fig4}b underscore the system's adaptive shift in operating mode as dictated by the CR functionality of the STAR-RIS, which reallocates resources in response to rising QoS demands.

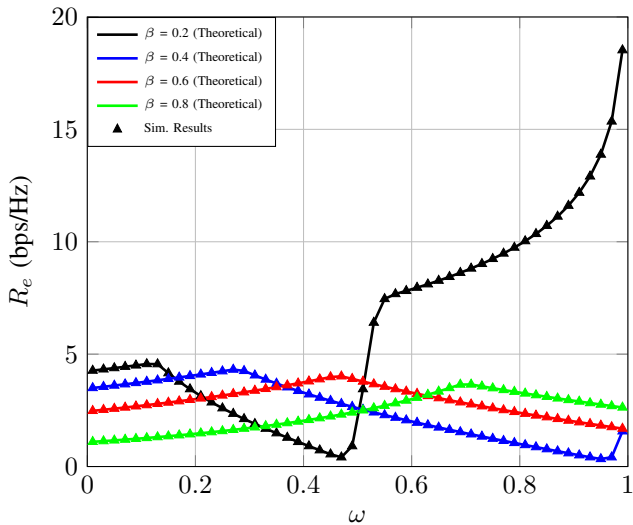
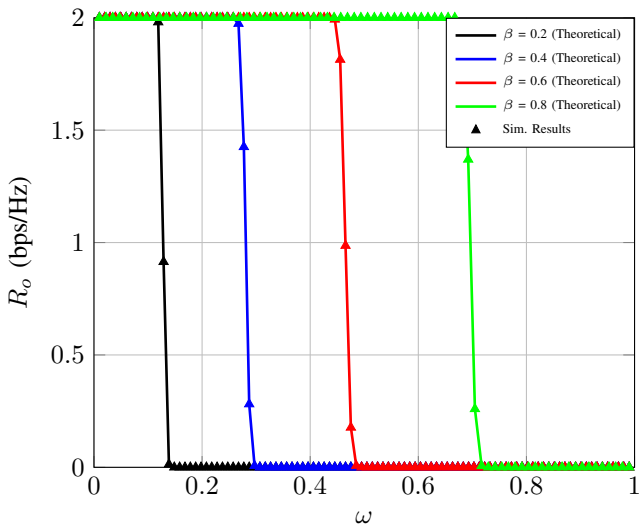
\begin{figure}[h!]
\centering
\begin{subfigure}{\linewidth}
\centering
\begin{tikzpicture}
\begin{axis}[width=0.99\linewidth,
xlabel = {$\omega$},
ylabel = {$R_e$ (bps/Hz)},
xmin = 0, xmax = 1,
ymin = 0, ymax = 20,
xtick = {0,0.2,...,1},
ytick = {0,5,...,20},
grid = major, legend image post style={xscale=0.9, every mark/.append style={solid}},
legend cell align = {left},
legend style={at={(0,1)}, anchor=north west, font = \tiny} ]
\addplot[
black,
no marks,
line width = 1pt,
solid]
table{Figures/omega/ER/beta0.2_cf.txt}; \addlegendentry{$\beta$ = 0.2 (Theoretical)}
\addplot[
blue,
no marks,
line width = 1pt,
solid]
table{Figures/omega/ER/beta0.4_cf.txt}; \addlegendentry{$\beta$ = 0.4 (Theoretical)}
\addplot[
red,
no marks,
line width = 1pt,
solid]
table{Figures/omega/ER/beta0.6_cf.txt}; \addlegendentry{$\beta$ = 0.6 (Theoretical) }
\addplot[
green,
no marks,
line width = 1pt,
solid]
table{Figures/omega/ER/beta0.8_cf.txt}; \addlegendentry{$\beta$ = 0.8 (Theoretical) }
\addplot[
black,
only marks,
mark=triangle*,
mark size = 2,]
table{Figures/omega/ER/beta0.2_mc.txt}; 
\addlegendentry{Sim. Results}
\addplot[
blue,
only marks,
mark=triangle*,
mark size = 2,]
table{Figures/omega/ER/beta0.4_mc.txt}; 
\addplot[
red,
only marks,
mark=triangle*,
mark size = 2,]
table{Figures/omega/ER/beta0.6_mc.txt}; 
\addplot[
green,
only marks,
mark=triangle*,
mark size = 2,]
table{Figures/omega/ER/beta0.8_mc.txt}; 
\end{axis}
\end{tikzpicture}
\subcaption{Ergodic rate of the SU.}
\end{subfigure}

\begin{subfigure}{\linewidth}
\centering
\begin{tikzpicture}
\begin{axis}[width=0.99\linewidth,
xlabel = {$\omega$},
ylabel = {$R_o$ (bps/Hz)},
xmin = 0, xmax = 1,
ymin = 0, ymax = 2,
xtick = {0,0.2,...,1},
grid = major, legend image post style={xscale=0.9, every mark/.append style={solid}},
legend cell align = {left},
legend style={at={(1,1)}, anchor=north east, font = \tiny} ]
\addplot[
black,
no marks,
line width = 1pt,
solid]
table{Figures/omega/OR/beta0.2_cf.txt}; \addlegendentry{$\beta$ = 0.2 (Theoretical)}
\addplot[
blue,
no marks,
line width = 1pt,
solid]
table{Figures/omega/OR/beta0.4_cf.txt}; \addlegendentry{$\beta$ = 0.4 (Theoretical)}
\addplot[
red,
no marks,
line width = 1pt,
solid]
table{Figures/omega/OR/beta0.6_cf.txt}; \addlegendentry{$\beta$ = 0.6 (Theoretical)}

\addplot[
green,
no marks,
line width = 1pt,
solid]
table{Figures/omega/OR/beta0.8_cf.txt}; \addlegendentry{$\beta$ = 0.8 (Theoretical)}
\addplot[
black,
only marks,
mark=triangle*,
mark size = 2,]
table{Figures/omega/OR/beta0.2_mc.txt}; \addlegendentry{Sim. Results}
\addplot[
blue,
only marks,
mark=triangle*,
mark size = 2,]
table{Figures/omega/OR/beta0.4_mc.txt};
\addplot[
red,
only marks,
mark=triangle*,
mark size = 2,]
table{Figures/omega/OR/beta0.6_mc.txt};
\addplot[
green,
only marks,
mark=triangle*,
mark size = 2,]
table{Figures/omega/OR/beta0.8_mc.txt};
\end{axis}
\end{tikzpicture}
\caption{Outage rate of the PU.}
\end{subfigure}
    \caption{Ergodic rate and outage rate versus $\omega$ for various $\beta$ values.}
    \label{fig7}
\end{figure}

Finally, Figs. \ref{fig7}a and \ref{fig7}b illustrate the influence of the spatial deployment parameter $\omega$ on the performance of the proposed STAR-RIS-assisted CR-RSMA system, evaluated for different values of the element allocation parameter $\beta$. Specifically, Fig. \ref{fig7}a shows the ergodic rate of the SU, while Fig. \ref{fig7}b presents the corresponding outage rate of the PU. As it can be seen, the theoretical expressions remain consistent with the simulation results, clearly reflecting the sharp transitions that characterize the system's behavior as $\omega$ varies. When $\beta$ is set to $0.2$, the STAR-RIS is predominantly configured for transmission, and for small values of $\omega$, where the RIS is located near the PU, both users can be simultaneously supported. As $\omega$ increases, the path loss associated with the reflected signal also increases, which weakens the PU’s effective channel gain and causes the outage rate in Fig. \ref{fig7}b to rise abruptly once reliable PU support is no longer feasible. Moreover, this loss in PU coverage affects the SU, whose ergodic rate in Fig. \ref{fig7}a, initially constrained by interference, increases sharply once the system reallocates all resources to the SU after the PU experiences outage. 
\textcolor{black}{This finding is also consistent with the PU's SINR expression, where the received power decreases as $\omega$
 increases.
}
Furthermore, as $\beta$ increases, a larger fraction of RIS elements is dedicated to reflection, enhancing the ability of the PU to maintain service at higher $\omega$ values. However, this comes at the cost of the SU performance, which remains limited even after the PU is no longer supported. Thus, Figs. \ref{fig7}a and \ref{fig7}b further highlight the unique adaptability of the CR functionality of the STAR-RIS, which can support high-quality CR services based on the PU and the SU distances and the channel conditions.

\section{Conclusions}
In this paper, we introduced the CR functionality as a tailored electromagnetic configuration for STAR-RIS-assisted CR-RSMA systems, enabling adaptive transmission and reflection control through the joint application of element and power splitting. Specifically, we demonstrated that the proposed functionality can dynamically adapt the STAR-RIS configuration in response to user priorities and interference conditions, supporting the stringent QoS requirements of the PU while enhancing the long-term performance of the secondary user. To characterize the system behavior, we derived closed-form SINR expressions and obtained analytical results for the ergodic rate of the secondary user and the outage rate of the PU using Gamma-based approximations, and we developed a case-based performance framework that captures the impact of varying channel conditions and STAR-RIS configurations on achievable rates. The provided simulation results confirmed the accuracy of the theoretical analysis and revealed critical trade-offs between STAR-RIS element allocation and RS parameters, highlighting the flexibility of the proposed approach in balancing performance and interference control. As a result, the presented findings establish the CR functionality as a practical and efficient solution for supporting future CR-RSMA networks empowered by PWEs.

\appendices
\section{\textsc{Proof of Proposition 3}}
\label{case2append}
Similarly with $\text{ER}_1$, the ergodic rate of the SU in Case II based on \eqref{su_rates} is given by
\begin{equation}
\small
\text{ER}_2 = \int_0^{\infty} \mathcal{I}_{\mathrm{in}}f_Y(y) dy,
\label{er2_integral}
\end{equation}
where
\begin{equation}
\small
\mathcal{I}_{\mathrm{in}} = \int_{\hat{\gamma}_p /L_p}^{\frac{\hat{\gamma}_p}{L_p}((1 - \alpha)L_s y+1)} \log_2\left(\frac{(1+\zeta_2)L_p\, x}{\hat{\gamma}_p}-\zeta_2\right) f_X(x)dx.
\end{equation}
According to the condition of Case II, $X$ and $Y$ follow a Gamma distribution with shape parameter $k$ and scale parameter $\theta_{e}$. Thus, by setting $t = \frac{(1+\zeta_2)}{\hat{\gamma}_p}y - \zeta_2$,  $\tau_1={\hat{\gamma}_p}/{L_p\theta_{e}}$, $\tau_2 = \tau_1/(\zeta_2+1)$ and converting $\log_2(\cdot)$ into natural logarithm form, we obtain
\begin{equation}
\small
\mathcal{I_{\mathrm{in}}} = \frac{e^{-\frac{\zeta_2\, \tau_1}{(\zeta_2 + 1)}}{\hat{\gamma}_p^k}\int_1^{(\zeta_2+1)(1-\alpha )L_s x+1} \ln(t) (\zeta_2+t)^{k-1}e^{-\frac{\tau_1\, t}{(\zeta_2+1)}} \, dt}{\Gamma(k) \, \theta_{e}^k\, \ln 2      ((\zeta_2 + 1) L_p)^k}.
\end{equation}
Additionally, by expanding the term $(\zeta_2+t)^{k-1}$ through binomial series
, and performing some algebraic manipulations, $\mathcal{I}_{in}$ can be rewritten as 
\begin{equation}
\small
\begin{split}
&\mathcal{I}_\mathrm{in}=\frac{e^{-\frac{\zeta_2\, \tau_1}{(\zeta_2 + 1)}}\tau_2^k}{\Gamma(k)\ln 2}\sum_{i_2=0}^{k-1}\hspace{-0.12cm}\binom{k-1}{i_2}\,\zeta_2^{i_2} \int_1^{(\zeta_2+1)(1-\alpha )x+1} \hspace{-1.4cm}\ln(t)\, t^{k-1-i_2} e^{-\frac{\tau_1 \,t}{\zeta_2+1}} \,dt.
\label{prof:in11}
\end{split}
\end{equation}
Moreover, by setting $z = \frac{\tau_1 \, t}{\zeta_2+1}$ and performing some algebraic manipulations, \eqref{prof:in11} can be reformulated as
\begin{equation}
\small
\begin{split}
&\mathcal{I}_\mathrm{in} = \frac{e^{-\frac{\zeta_2\, \tau_1}{\zeta_2 + 1}}}{\Gamma(k)\ln 2}
\sum_{i_2=0}^{k-1} \binom{k-1}{i_2}
\left({\tau_1\,\zeta_2}\right)^{i_2} \\
&\times \left[ 
\int_{0}^{\tau_2+\psi x} \ln(t)\, t^{k-i_2-1}e^{-t}\,dt -
\int_{0}^{\tau_2} \ln(t)\, t^{k-i_2-1}e^{-t}\,dt \right. \\
& + \ln\left(\frac{1}{\tau_2}\right)
\left(\int_{0}^{\tau_2+\psi x} t^{k-i_2-1}e^{-t}\,dt -
\int_{0}^{\tau_2} t^{k-i_2-1}e^{-t}\,dt\right)
\left. \vphantom{\int_{0}^{\tau_2+\psi x}} \right].
\end{split}
\label{prof:in12}
\end{equation}
Thus, by setting $\kappa=k-i_2$ and leveraging the definition of lower incomplete gamma function $\gamma(k, z)$ along with \cite[Eq. (1.6.10.1)]{Prudnikov}, ~\eqref{prof:in12} can be expressed as
\begin{equation}
\small
\begin{split}
\mathcal{I}_\mathrm{in} &= \frac{e^{-\frac{\zeta_2\, \tau_1}{\zeta_2 + 1}}}{\Gamma(k)\ln 2}
\sum_{i_2=0}^{k-1} \binom{k-1}{i_2}
\left({\tau_1\,\zeta_2}\right)^{i_2}\,\mathcal{I}(x)
\end{split}
\label{prof:in13}
\end{equation}
where ${}_2F_2(\cdot)$ is the generalized hypergeometric function, and $\mathcal{I}(x)$ is equal to
\begin{equation}
\small
\begin{split}
\mathcal{I}(x)=\; & \gamma(\kappa,\tau_2+\psi x) \left[\ln(\tau_2+\psi x) + \ln \bigl({1}/{\tau_2}\bigr)\right] \\[1ex]
&- {}_2F_2(\kappa,\kappa;\kappa+1,\kappa+1;-(\tau_2+\psi x))\,\frac{\bigl(\tau_2+\psi x\bigr)^\kappa}{\kappa^2} 
\\[1ex]&+{}_2F_2(\kappa,\kappa;\kappa+1,\kappa+1;-\tau_2)\,\frac{\tau_2^\kappa}{\kappa^2}
\\[1ex]
&- \gamma(\kappa,\tau_2) \left[\ln(\tau_2) + \ln\bigl(1/\tau_2\bigr)\right]\,.
\end{split}
\end{equation}
Thus, by substituting~\eqref{prof:in13} into~\eqref{er2_integral}, $\text{ER}_2$ can be rewritten as  
\begin{equation}
\small
\begin{split}
&\text{ER}_2 = \frac{e^{-\frac{\zeta_2\, \tau_1}{\zeta_2 + 1}}}{\Gamma(k)^2\, \theta_{e}^k\,\ln 2}
\sum_{i_2=0}^{k-1} \binom{k-1}{i_2}
\left({\tau_1\,\zeta_2}\right)^{i_2} \\&
\times \int_0^\infty \mathcal{I}(x) e^{-x/\theta_{e}} x^{k-1}\, dx.
\label{prof:case2}
\end{split}
\end{equation}
To facilitate the derivation of \eqref{er:c2}, we proceed by rewriting \eqref{prof:case2} as
\begin{equation}
\small
\begin{split}
&\text{ER}_2 = \frac{e^{-\frac{\zeta_2\, \tau_1}{\zeta_2 + 1}}}{\Gamma(k)^2\, \theta_{e}^k\,\ln 2}
\sum_{i_2=0}^{k-1} \binom{k-1}{i_2}
\left({\tau_1\,\zeta_2}\right)^{i_2} \\&
\times \left( \mathcal{I}_1 + \mathcal{I}_2-\mathcal{I}_3+\mathcal{I}_4 \right),
\label{prof:case22}
\end{split}
\end{equation}
where $\mathcal{I}_1$, $\mathcal{I}_2$, $\mathcal{I}_3$ and $\mathcal{I}_4$ are given as \begin{equation}
\small
\mathcal{I}_1 = \int_0^\infty \ln(\tau_2 + \psi x)\, \gamma(\kappa, \tau_2 + \psi x)\, x^{k-1}e^{-x/\theta_{e}}\, dx,
\label{prof:1}
\end{equation}
\begin{equation}
\small
\mathcal{I}_2 = \ln(1/\tau_2) \int_0^\infty \gamma(\kappa, \tau_2 + \psi x)\, x^{k-1}e^{-x/\theta_{e}}\, dx,
\label{prof:2}
\end{equation}
\begin{equation}
\small
\begin{split}
\mathcal{I}_3 =\hspace{-0.1cm} \int_0^\infty &\frac{(\tau_2 + \psi x)^\kappa}{\kappa^2}\, {}_2F_2(\kappa, \kappa; \kappa + 1, \kappa + 1; -(\tau_2 + \psi x))\, \\& \times x^{k-1}e^{-x/\theta_{e}}\, dx,
\end{split}
\label{prof:3}
\end{equation}
and
\begin{equation}
\small
\begin{split}
&\mathcal{I}_4 =\hspace{-0.1cm} \int_0^\infty\Big( {}_2F_2(\kappa,\kappa;\kappa+1,\kappa+1;-\tau_2)\,\frac{\tau_2^\kappa}{\kappa^2}
\\&- \gamma(\kappa,\tau_2) \left[\ln(\tau_2) + \ln\bigl(1/\tau_2\bigr)\right]\Big)\, x^{k-1}e^{-x/\theta_{e}}\, dx,
\label{prof:q}
\end{split}
\end{equation}

Therefore, to conclude the proof, we need to explicitly solve the integrals given by \eqref{prof:1}, \eqref{prof:2}, \eqref{prof:3}, and \eqref{prof:q}.
\subsubsection{$\mathcal{I}_1$} To derive $\mathcal{I}_{1}$ we can express the lower incomplete gamma function as $\gamma(\kappa, z) = \Gamma(\kappa) - \Gamma(\kappa, z)$, thereby reformulating~\eqref{prof:1} as
\begin{equation}
\small
\mathcal{I}_{1} = \mathcal{I}_{1}^{(1)} - \mathcal{I}_{1}^{(2)},
\label{prof:i1_final}
\end{equation}
where $\mathcal{I}_1^{(1)} = \int_0^{\infty} \ln(\tau_2+\psi x) \, \Gamma(\kappa)\,x^{k-1}\,e^{-x/\theta_{e}} \, dx$, and $\mathcal{I}_1^{(2)}=\int_0^{\infty} \ln(\tau_2+\psi x) \, \Gamma(\kappa, z)\, x^{k-1}\,e^{-x/\theta_{e}} \, dx$. By taking into account the integral $\mathcal{I}_1^{(1)}$, through algebraic manipulations, it can be rewritten as
\begin{equation}
\small
\begin{split}
\mathcal{I}_1^{(1)} &= \ln (\tau_2)\, \Gamma (k)\,\theta_{e}^k \, \Gamma (\kappa)\\
& \quad +\theta_{e}^k \,
\Gamma (\kappa) \int_0^{\infty} \ln \left(1+\frac{\psi\, \theta_{e} x}{\tau_2} \right) x^{k-1}\,e^{-x} \,dx,
\label{prof:i1}
\end{split}
\end{equation}
and by utilizing ~\cite[Eq.~(4.337.5)]{Ryzhik2014}, \eqref{prof:i1} can be expressed as in \eqref{prof:i11_cf}. 
\begin{figure*}[t]
\begin{equation}
  \small
  \begin{split}
  \mathcal{I}_1^{(1)} = \ln(\tau_2)\,\Gamma(k)\,\theta_{e}^k\, \Gamma(\kappa) + \theta_{e}^k \,\Gamma(\kappa) \sum_{q_2=0}^{k-1} \frac{(k-1)!}{(k-q_2-1)!} \, \times \left[\frac{(-1)^{k-q_2-2}e^{\frac{\tau_2}{\psi\,\theta_{e}}} \operatorname{Ei}\left(-\frac{\tau_2}{\psi\,\theta_{e}}\right) }{\left(\frac{\psi\,\theta_{e}}{\tau_2}\right)^{k-q_2-1}} +\sum_{t_2=1}^{k-q_2-1}(t_2-1)!\left(-\frac{\tau_2}{\psi\,\theta_{e}}\right)^{k-t_2-q_2-1}\right]
  \end{split}
  \label{prof:i11_cf}
  \end{equation}
  \vspace{1ex}\hrule 
\end{figure*}
In addition, to calculate $\mathcal{I}_1^{(2)}$ we utilize the series expansion of upper incomplete Gamma function, thus expressing $\mathcal{I}_1^{(2)}$ as
\begin{equation}
\small
\begin{split}
\mathcal{I}_1^{(2)} & = \sum_{a_2=0}^{\kappa-1}\int_0^{\infty}\hspace{-0.2cm} \ln (\tau_2 + \psi x) e^{-(\tau_2+ \psi x)}(\tau_2 + \psi x)^{a_2} \, x^{k-1} \,e^{-x/\theta_{e}}\, dx \\
& \quad \times\frac{(\kappa-1)!}{a_2!} ,
\end{split}
\label{prof:i2}
\end{equation}
Moreover, by expanding the term $(\tau_2+\psi x)^{a_2}$ through binomial series and utilizing \eqref{pdf} and \cite[Eq.~(4.337.5)]{Ryzhik2014}, \eqref{prof:i2} can be rewritten as in \eqref{prof:i12_cf}, which concludes the derivation of $\mathcal{I}_1$. 
\begin{figure*}[t]
  \begin{equation}
\small
  \begin{split}
  &\mathcal{I}_1^{(2)} = \sum _{a_2=0}^{\kappa-1}\sum_{n_2}^{a_2}\binom{a_2}{n_2}\frac{(\kappa-1)! e^{-\tau_2}\,{\tau_2}^{a_2-n_2} \, \psi^{n_2}}{a_2!}\,\left(\frac{\theta_{e}}{\psi\,\theta_{e}+1}\right)^{k+{n_2}} \hspace{-0.1cm}\left\{\ln (\tau_2)\, \Gamma(k+n_2) + \left[\sum_{e_2=0}^{k+{n_2}-1} \frac{(k+{n_2}-1)!}{(k+{n_2}-e_2-1)!} \right. \right. \\[1ex]
  &\hspace{-0.15cm}\left. \left.
  \times \hspace{-0.05cm} \left(\frac{(-1)^{k+{n_2}-{e_2}-2}}{\left(\frac{\psi\, \theta_{e}}{\tau_2 (\psi\, \theta_{e} + 1)}\right)^{k+{n_2}-e_2-1}}e^{\frac{\tau_2 (\psi\, \theta_{e} + 1)}{\psi\, \theta_{e}}} \operatorname{Ei}\left(-\frac{\tau_2 (\psi\, \theta_{e} + 1)}{\psi\, \theta_{e}}\right)  +
  \sum_{r_2=1}^{k+{n_2}-e_2-1} (r_2-1)! \left(-\frac{\tau_2 (\psi\, \theta_{e} + 1)}{\psi\, \theta_{e}}\right)^{k+{n_2}-{e_2}-{r_2}-1}  \right)\right]\right\}
  \end{split}
  \label{prof:i12_cf}
  \end{equation}
  \vspace{1ex}\hrule 
\end{figure*}
\subsubsection{$\mathcal{I}_2$} Similarly with $\mathcal{I}_1$, by using $\gamma(\kappa, z) = \Gamma(\kappa) - \Gamma(\kappa, z)$ and expanding the upper incomplete gamma function through its binomial series form, \eqref{prof:2} can be written as  
\begin{equation}
\small
\begin{split}
\mathcal{I}_2 &= \ln(1/\tau_2)\,\Gamma(k)\,\theta_{e}^k \,\Gamma(\kappa)\hspace{-0.07cm}- \sum _{b_2=0}^{\kappa-1} \frac{(\kappa-1)!}{{b_2}!}\,\ln(1/\tau_2)e^{-\tau_2} \\& \times \int_0^{\infty }(\tau_2+\psi\,x)^{b_2} x^{k-1} \,e^{-\frac{(\psi \,\theta_{e}+1)}{\theta_{e}}} \, dx.
\end{split}
\end{equation}
Finally, by expanding the term $(\tau_2+\psi\,x)^{b_2}$ through binomial series, after some algebraic manipulation, $\mathcal{I}_2$ can be derived as
\begin{equation}
\small
\begin{split}
\mathcal{I}_2 &= \ln(1/\tau_2)\,\Gamma(k)\,\theta_{e}^k \,\Gamma(\kappa)-\left(\sum_{b_2=0}^{\kappa-1} \sum_{c_2=0}^{b_2}\binom{b_2}{c_2}\frac{(\kappa-1)!}{b_2!} \right.
\\&\left. \times\ln(1/\tau_2) \,\Gamma (k+c_2)\, \tau_2^{b_2-c_2}\, \psi^{c_2}\,e^{-\tau_2}\, \left(\frac{\theta_{e}}{(\psi\, \theta_{e}+1)}\right)^{k+c_2}\right).
\label{}
\end{split}
\end{equation}
\subsubsection{$\mathcal{I}_3$} To calculate $\mathcal{I}_3$, we use ${}_2F_2(\cdot)$ as ${}_2F_2(a_1, a_2; b_1, b_2; x) = \sum_{k = 0}^\infty \frac{(a_1)_k (a_2)_k}{(b_1)_k (b_2)_k} \frac{x^k}{k!}$ and rewrite \eqref{prof:3} as
\begin{equation}
\small
\mathcal{I}_3= \sum_{j_2 = 0}^{\infty}\int_{0}^{\infty} \frac{ (\kappa)_{j_2}(\kappa)_{j_2}(\tau_2+\psi x)^{\kappa+j_2}}{\kappa^2(\kappa+1)_{j_2}(\kappa+1)_{j_2}} x^{k-1}\,e^{-x/\theta_{e}}\, dx.
\label{equati3}
\end{equation}
Moreover, by taking into account that $(a_1)_k=\frac{\Gamma(a+n)}{\Gamma(a)}$ and expanding the term $(\tau_2+\psi\, x)^{\kappa+j_2}$ through the binomial series, and after some algebraic manipulations, \eqref{equati3} can be rewritten as 
\begin{equation}
\small
\begin{split}
\mathcal{I}_3 &= \sum_{j_2=0}^{\infty}\sum_{f_2=0}^{\kappa+j_2}\binom{\kappa+j_2}{f_2}\hspace{-0.14cm}  \left(\frac{1}{\kappa+j_2}\right)^2\frac{(-1)^{j_2}}{{j_2}!} \psi^f \theta_{e}^{k+f} \tau_2^{\kappa+j_2 - f_2}\\
& \quad \times \Gamma (k+f_2),
\label{equat-i3-cf}
\end{split}
\end{equation}
which concludes the derivation of $\mathcal{I}_3$.
\subsubsection{$\mathcal{I}_4$} To obtain $\mathcal{I}_4$, we utilize the integral definition of gamma function and thus, after some algebraic manipulations, \eqref{prof:q} can be rewritten as 
\begin{equation}
\small
\begin{split}
\mathcal{I}_4 & ={}_2F_2(\kappa,\kappa;\kappa+1,\kappa+1;-\tau_2)\,\frac{\tau_2^\kappa}{\kappa^2}\,\theta_{e}^k\,\Gamma(k)
\\& \quad - \gamma(\kappa,\tau_2) \left[\ln(\tau_2) + \ln\bigl(1/\tau_2\bigr)\right]\,\theta_{e}^k\,\Gamma(k),
\label{}
\end{split}
\end{equation}
which concludes the derivation of $\mathcal{I}_4$.

After obtaining $\mathcal{I}_1$, $\mathcal{I}_2$, $\mathcal{I}_3$, and $\mathcal{I}_44$ we substitute them into \eqref{prof:case22} which leads to~\eqref{er:c2}, thus concluding the proof.

\bibliographystyle{IEEEtran}
\bibliography{Reference}

\begin{thebibliography}{10}
\providecommand{\url}[1]{#1}
\csname url@samestyle\endcsname
\providecommand{\newblock}{\relax}
\providecommand{\bibinfo}[2]{#2}
\providecommand{\BIBentrySTDinterwordspacing}{\spaceskip=0pt\relax}
\providecommand{\BIBentryALTinterwordstretchfactor}{4}
\providecommand{\BIBentryALTinterwordspacing}{\spaceskip=\fontdimen2\font plus
\BIBentryALTinterwordstretchfactor\fontdimen3\font minus \fontdimen4\font\relax}
\providecommand{\BIBforeignlanguage}[2]{{%
\expandafter\ifx\csname l@#1\endcsname\relax
\typeout{** WARNING: IEEEtran.bst: No hyphenation pattern has been}%
\typeout{** loaded for the language `#1'. Using the pattern for}%
\typeout{** the default language instead.}%
\else
\language=\csname l@#1\endcsname
\fi
#2}}
\providecommand{\BIBdecl}{\relax}
\BIBdecl

\bibitem{DINGSurvey2017}
Z.~Ding, X.~Lei, G.~K. Karagiannidis, R.~Schober, J.~Yuan, and V.~K. Bhargava, ``A survey on non-orthogonal multiple access for {5G} networks: {Research} challenges and future trends,'' \emph{IEEE Trans. Veh. Technol.}, vol.~35, no.~10, pp. 2181--2195, 2017.

\bibitem{Mishra2022RSMA}
A.~Mishra, Y.~Mao, O.~Dizdar, and B.~Clerckx, ``Rate-splitting multiple access for {6G—Part I: Principles}, applications and future works,'' \emph{IEEE Commun. Lett.}, vol.~26, no.~10, pp. 2232--2236, 2022.

\bibitem{LiU2017NOMA}
Y.~Liu, Z.~Qin, M.~Elkashlan, Y.~Gao, and L.~Hanzo, ``Enhancing the physical layer security of non-orthogonal multiple access in large-scale networks,'' \emph{IEEE Trans. Wirel. Commun.}, vol.~16, no.~3, pp. 1656--1672, 2017.

\bibitem{proceed}
N.~G. Evgenidis, N.~A. Mitsiou, V.~I. Koutsioumpa, S.~A. Tegos, P.~D. Diamantoulakis, and G.~K. Karagiannidis, ``{Multiple Access in the Era of Distributed Computing and Edge Intelligence},'' \emph{Proc. IEEE}, vol. 112, no.~9, pp. 1497--1526, 2024.

\bibitem{Tegos2022}
S.~A. Tegos, P.~D. Diamantoulakis, and G.~K. Karagiannidis, ``{On the performance of uplink rate-splitting multiple access},'' \emph{IEEE Commun. Lett.}, vol.~26, no.~3, pp. 523--527, Mar. 2022.

\bibitem{LiuCRNOMA2022}
H.~Liu, Z.~Bai, H.~Lei, G.~Pan, K.~J. Kim, and T.~A. Tsiftsis, ``A new rate splitting strategy for uplink {CR-NOMA} systems,'' \emph{IEEE Trans. Veh. Technol.}, vol.~71, no.~7, pp. 7947--7951, 2022.

\bibitem{YUE2023CR-RSMA}
Y.~Xiao, S.~A. Tegos, P.~D. Diamantoulakis, Z.~Ma, and G.~K. Karagiannidis, ``{On the ergodic rate of cognitive radio inspired uplink multiple access},'' \emph{IEEE Commun. Lett.}, vol.~27, no.~1, pp. 95--99, 2023.

\bibitem{Munochiveyi2021}
M.~Munochiveyi, A.~C. Pogaku, D.-T. Do, A.-T. Le, M.~Voznak, and N.~D. Nguyen, ``Reconfigurable intelligent surface aided multi-user communications: {State}-of-the-art techniques and open issues,'' \emph{IEEE Access}, vol.~9, pp. 118\,584--118\,605, 2021.

\bibitem{liaskosmagazine}
C.~Liaskos, S.~Nie, A.~Tsioliaridou, A.~Pitsillides, S.~Ioannidis, and I.~Akyildiz, ``A new wireless communication paradigm through software-controlled metasurfaces,'' \emph{IEEE Commun. Mag.}, vol.~56, no.~9, pp. 162--169, 2018.

\bibitem{direnzo2022}
M.~Di~Renzo, F.~H. Danufane, and S.~Tretyakov, ``Communication models for reconfigurable intelligent surfaces: From surface electromagnetics to wireless networks optimization,'' \emph{Proceedings of the IEEE}, vol. 110, no.~9, pp. 1164--1209, 2022.

\bibitem{tegos2022distribution}
S.~A. Tegos, D.~Tyrovolas, P.~D. Diamantoulakis, C.~K. Liaskos, and G.~K. Karagiannidis, ``On the distribution of the sum of double-{Nakagami}-$ m $ random vectors and application in randomly reconfigurable surfaces,'' \emph{IEEE Trans. Veh. Technol.}, vol.~71, no.~7, pp. 7297--7307, 2022.

\bibitem{yuanwei2021RIS}
Y.~Liu, X.~Liu, X.~Mu, T.~Hou, J.~Xu, M.~Di~Renzo, and N.~Al-Dhahir, ``Reconfigurable intelligent surfaces: {Principles} and opportunities,'' \emph{IEEE Commun. Surv. Tutor.}, vol.~23, no.~3, pp. 1546--1577, 2021.

\bibitem{liu2021star}
Y.~Liu, X.~Mu, J.~Xu, R.~Schober, Y.~Hao, H.~V. Poor, and L.~Hanzo, ``{STAR}: Simultaneous transmission and reflection for 360° coverage by intelligent surfaces,'' \emph{IEEE Wirel. Commun.}, vol.~28, no.~6, pp. 102--109, 2021.

\bibitem{zeris}
D.~Tyrovolas, S.~A. Tegos, V.~K. Papanikolaou, Y.~Xiao, P.-V. Mekikis, P.~D. Diamantoulakis, S.~Ioannidis, C.~K. Liaskos, and G.~K. Karagiannidis, ``Zero-energy reconfigurable intelligent surfaces {(zeRIS)},'' \emph{IEEE Trans. Wirel. Commun.}, vol.~23, no.~7, pp. 7013--7026, 2024.

\bibitem{leris}
D.~Tyrovolas, D.~Bozanis, S.~A. Tegos, V.~K. Papanikolaou, P.~D. Diamantoulakis, C.~K. Liaskos, R.~Schober, and G.~K. Karagiannidis, ``Empowering programmable wireless environments with optical anchor-based positioning,'' \emph{IEEE Netw.}, vol.~39, no.~1, pp. 14--20, 2025.

\bibitem{activeris}
R.~Long, Y.-C. Liang, Y.~Pei, and E.~G. Larsson, ``Active reconfigurable intelligent surface-aided wireless communications,'' \emph{IEEE Trans. Wirel. Commun.}, vol.~20, no.~8, pp. 4962--4975, 2021.

\bibitem{YuanJie2021}
J.~Yuan, Y.-C. Liang, J.~Joung, G.~Feng, and E.~G. Larsson, ``Intelligent reflecting surface-assisted cognitive radio system,'' \emph{IEEE Trans. Commun.}, vol.~69, no.~1, pp. 675--687, 2021.

\bibitem{Vu2022}
T.-H. Vu, T.-V. Nguyen, D.~B.~d. Costa, and S.~Kim, ``Reconfigurable intelligent surface-aided cognitive noma networks: Performance analysis and deep learning evaluation,'' \emph{IEEE Trans. Wirel. Commun.}, vol.~21, no.~12, pp. 10\,662--10\,677, 2022.

\bibitem{Weiheng2022}
W.~Jiang, Y.~Zhang, J.~Zhao, Z.~Xiong, and Z.~Ding, ``Joint transmit precoding and reflect beamforming design for {IRS}-assisted mimo cognitive radio systems,'' \emph{IEEE Trans. Wirel. Commun.}, vol.~21, no.~6, pp. 3617--3631, 2022.

\bibitem{xia2024shortpacket}
C.~Xia, Z.~Xiang, H.~Liu, and J.~Meng, ``Reliability performance of {RIS-CR-NOMA} based short-packet communications,'' in \emph{2024 IEEE 16th International Conference on Advanced Infocomm Technology (ICAIT)}, 2024, pp. 60--67.

\bibitem{Lv2025MEC}
L.~Lv, H.~Luo, L.~Yang, Z.~Ding, A.~Nallanathan, N.~Al-Dhahir, and J.~Chen, ``Ris-assisted wireless powered mec: Multiple access design and resource allocation,'' \emph{IEEE Trans. Wirel. Commun.}, vol.~24, no.~2, pp. 984--1000, 2025.

\bibitem{RSMATsiftsis2023}
\BIBentryALTinterwordspacing
P.~Liu, G.~Jing, H.~Liu, L.~Yang, and T.~A. Tsiftsis, ``Intelligent reflecting surface-assisted cognitive radio-inspired rate-splitting multiple access systems,'' \emph{Digit. Commun. Netw.}, vol.~9, no.~3, pp. 655--666, 2023. [Online]. Available: \url{https://www.sciencedirect.com/science/article/pii/S2352864822002413}
\BIBentrySTDinterwordspacing

\bibitem{li2024star}
H.~Li, Y.~Liu, X.~Mu, Y.~Chen, P.~Zhiwen, and X.~You, ``{STAR-RIS in cognitive radio networks},'' \emph{IEEE Trans. Wireless Commun.}, 2024.

\bibitem{Wang2023STAROutage}
S.~Wang, B.~Lian, X.~Gao, X.~Li, and M.~Zeng, ``Outage performance analysis of {STAR-RIS} assisted {CR-NOMA} networks,'' in \emph{GLOBECOM 2023 - 2023 IEEE Global Communications Conference}, 2023, pp. 4411--4417.

\bibitem{Security2024STAR}
X.~Li, J.~Zhang, C.~Han, W.~Hao, M.~Zeng, Z.~Zhu, and H.~Wang, ``Reliability and security of cr-star-ris-noma-assisted iot networks,'' \emph{IEEE Internet Things J.}, vol.~11, no.~17, pp. 27\,969--27\,980, 2024.

\bibitem{IndustrySTAR2024}
X.~Li, X.~Gao, L.~Yang, H.~Liu, J.~Wang, and K.~M. Rabie, ``{Performance Analysis of STAR-RIS-CR-NOMA-Based Consumer IoT Networks for Resilient Industry 5.0},'' \emph{IEEE Trans. Consum. Electron.}, vol.~70, no.~1, pp. 1380--1391, 2024.

\bibitem{10014691}
M.~Katwe, K.~Singh, B.~Clerckx, and C.-P. Li, ``Improved spectral efficiency in {STAR-RIS} aided uplink communication using rate splitting multiple access,'' \emph{IEEE Trans. Wirel. Commun.}, vol.~22, no.~8, pp. 5365--5382, 2023.

\bibitem{9755045}
H.~Liu, Z.~Bai, H.~Lei, G.~Pan, K.~J. Kim, and T.~A. Tsiftsis, ``A new rate splitting strategy for uplink {CR-NOMA} systems,'' \emph{IEEE Trans. Vehicular Tech.}, vol.~71, no.~7, pp. 7947--7951, 2022.

\bibitem{tyrharq}
D.~Tyrovolas, P.-V. Mekikis, S.~A. Tegos, P.~D. Diamantoulakis, C.~K. Liaskos, and G.~K. Karagiannidis, ``On the performance of {HARQ} in {IoT} networking with {UAV-mounted} reconfigurable intelligent surfaces,'' in \emph{2022 IEEE 95th Vehicular Technology Conference: (VTC2022-Spring)}, 2022, pp. 1--5.

\bibitem{HQAM}
T.~K. Oikonomou, D.~Tyrovolas, S.~A. Tegos, P.~D. Diamantoulakis, P.~Sarigiannidis, C.~Liaskos, and G.~K. Karagiannidis, ``On the performance of {RIS}-assisted networks with {HQAM},'' in \emph{2024 Joint European Conference on Networks and Communications \& 6G Summit (EuCNC/6G Summit)}, 2024, pp. 434--439.

\bibitem{Ryzhik2014}
I.~S. Gradshteyn and I.~M. Ryzhik, \emph{Table of integrals, series, and products}.\hskip 1em plus 0.5em minus 0.4em\relax Academic press, 2014.

\bibitem{Prudnikov}
A.~P. Prudnikov, Y.~A. Brychkov, and O.~I. Marichev, \emph{Integrals and Series. Volume 1: Elementary Functions}.\hskip 1em plus 0.5em minus 0.4em\relax Moscow: Taylor \& Francis, 1986.

\end{thebibliography}
\end{document}